\documentclass[tighten, onecolumn, trackchanges, 12pt]{aastex62}
\def\apjl{{ApJ}}                % Astrophysical Journal, Letters
\def\apjs{{ApJS}}               % Astrophysical Journal, Supplement
\def\ao{{Appl.~Opt.}}           % Applied Optics
\def\aap{{A\&A}}                % Astronomy and Astrophysics
\usepackage{listings}
\usepackage{color}
\usepackage{graphicx,times}
\usepackage{natbib}
\usepackage{amssymb}
\usepackage{newtxtext}
\usepackage[T1]{fontenc}
\usepackage{ae,aecompl}
\usepackage{threeparttable}
\usepackage{ulem}
\usepackage{txfonts}
\usepackage{amssymb}	% Extra maths symbols
\def\kms  {km~s$^{-1}$}

\def\kmsperkpc  {km~s$^{-1}$~kpc$^{-1}$}

\shorttitle{galactic geometry and kinematics traced by open clusters}
\shortauthors{He 2023}
\bibpunct{(}{)}{;}{a}{}{,}
\usepackage{hyperref}
\usepackage{longtable}
\usepackage{booktabs}
\definecolor{dkgreen}{rgb}{0,0.6,0}
\definecolor{gray}{rgb}{0.5,0.5,0.5}
\definecolor{mauve}{rgb}{0.58,0,0.82}
\definecolor{golden}{rgb}{0.86,0.65,0.01}

\usepackage{CJK}
\usepackage{graphicx}
\usepackage{hyperref}  
\usepackage{amssymb}  
\begin{document}
\begin{CJK*}{UTF8}{gbsn}
%
%\received{11 Jul~2022}
%\revised{15 Sep~2022}
%\accepted{          }
%
%
\title[]{~\textbf{Geometry and Kinematics of a Dancing Milky Way:\\ Unveiling the Precession and
Inclination Variation across the Galactic Plane via Open Clusters}}
\email{hezh@cwnu.edu.cn}
\author[0000-0002-6989-8192]{Zhihong He (何治宏)}
\affil{School of Physics and Astronomy, China West Normal University, No. 1 Shida Road, Nanchong 637002, China }
\vspace{10pt}

\begin{abstract}
This Letter presents a study of the geometry and motion of the Galactic disk using open clusters in the Gaia era. The findings suggest that the inclination $\theta_i$ of the Galactic disk increases gradually from the inner to the outer disk, with a shift in orientation at the Galactocentric radius of approximately 6~$\pm$~1~kpc. Furthermore, this study brings forth the revelation that the mid-plane of the Milky Way may not possess a stationary or fixed position. A plausible explanation is that the inclined orbits of celestial bodies within our Galaxy exhibit a consistent pattern of elliptical shapes, deviating from perfect circularity; however, more observations are needed to confirm this.
An analysis of the vertical motion along the Galactocentric radius reveals that the disk has warped with precession, and that the line-of-nodes shifts at different radii, aligning with the results from the classical Cepheids. Although there is uncertainty for precession/peculiar motion in Solar orbit, after considering the uncertainty, the study derives a median value of $\dot{\phi}_{LON}$ = 6.8~\kmsperkpc in the Galaxy. This value for the derived precession in the outer disk is lower than those in the literature due to the systematic motion in Solar orbit ($\theta_i$ = 0.6$^{\circ}$). The study also finds that the inclinational variation of the disk is significant and can cause systematic motion, with the variation rate $\dot{\theta}_i$ decreasing along the Galactic radius with a slope of -8.9~$\mu$as~yr$^{-1}$~kpc$^{-1}$. Moreover, the derived $\dot{\theta}_i$ in Solar orbit is 59.1~$\pm$~11.2$_{sample}$~$\pm$~7.7$_{V_{Z\odot}}$~$\mu$as~yr$^{-1}$, which makes it observable for high precision astrometry.
The all-sky open cluster catalog based on Gaia DR3 and Galactic precession/inclinational variation fits as well as Python code related to these fits are available at \url{https://nadc.china-vo.org/res/r101288/}.

\end{abstract}

\section{Introduction}\label{sec:intro}

When observing galaxies, astronomers often find that the disk is bent or tilted~\citep{Jarrett03} - this is known as the warp structure. It is believed that this phenomenon is caused by a variety of sources of gravitational influence, such as interactions with neighboring galaxies or dark matter halos~\citep[e.g.][]{Hunter69,Sparke88,Quinn93,Shen06}. The motion of stars and gas in a galaxy's disk can cause distortion in its shape by moving in different directions or orbits. To study the warp structure of galaxies, researchers use observations of the positions and movements of stars and gas clouds~\citep[e.g.][]{Sancisi76,Bosma81,Thilker05} along with simulations and models of galactic dynamics~\citep[e.g.][]{Pringle92,Velazquez99,Robin03,Rok10}. A deeper understanding of the warp structure is crucial to comprehend the evolution of galaxies and the underlying processes that drive their formation and transformation in the universe.

The warped structure in the outer disk of our Milky Way has been known for some time, first detected through the observation of HI gas~\citep{Kerr57,Burke57}. Studies have shown that the vertical angle increases to 3 degrees at a Galactocentric radius of around 16~kpc~\citep{Burton88}. Similarly, molecular CO clouds observed in the first and second Galactic quadrant have also demonstrated similar warp features in the outer arm~\citep{Dame11,Sun15}. However, the uncertainty of gas kinematic distances has made it difficult to study the geometric structure of the Galactic warp. It was only through the observation of classical Cepheids~\citep[][hereafter CCs]{Chen19,Skowron19} in recent years that a clear warp in the outer disk has become measurable. This allowed for the direct measurement of the Galactocentric radius (R$_{GC}$) and scale height of the warp. Findings indicate that both the gas and young star disk warp upwards in the first and/or second quadrants, and downwards in the third and/or fourth quadrants. Moreover, the absolute vertical amplitude reaches 0.3 to 5~kpc above the Galactic plane beyond R$_{GC}$ = 10 to 30~kpc ~\citep[e.g.][]{Gum60,Corredoira02,Nakanishi03, Levine06,Voskes06,Chen19,Skowron19,Skowron19b,Lemasle22}, highlighting the significant dimensions of the warp structure.

Furthermore, the orbital inclination of the warp can lead to a systematic velocity in the vertical direction, particularly close to the line-of-node (LON) where the warp surface intersects with the Galactic plane. During the ascending semicycle, this systematic velocity often displays an upward trend while the opposite is observed in the descending semicycle. Multiple statistical studies have identified this trend using various stellar tracers~\citep[][]{Corredoira14,Liu17,Poggio17}, especially with the improved accuracy of proper motion data from later Gaia releases~\citep{gaiadr2,gaiaedr3,gaiadr3}. Additionally, the vertical velocity of stars in low-latitude regions near the Galactic plane can be determined by its proper motion $\mu_b$ and is less influenced by the radial velocity~\citep[e.g.][]{Poggio18,Poggio20,Romero19,Wang20,Li20,Cheng20,Li23,Dehnen23}.

The question of whether the Milky Way's warped disk is precessing has sparked debate. ~\citet{Poggio20} and ~\citet{Cheng20} proposed that there is precession of around 10.9 and 13.6~\kmsperkpc in the outer Galactic disk, respectively, while others argue that there is no significant precession~\citep[e.g.][]{Wang20,Corredoira21}. The debate stems from the fact that the precession velocity calculation depends on different tracers and warp models. Recently, ~\citet{Dehnen23} introduced a new perspective on precession in the outer disk of the Milky Way based on classical Cepheids, which takes into account the unsteady inclination (known as the inclinational variation rate $\dot{\theta}_i$ in this context) for the first time. This study showed a precession rate that gradually decreases from 12 (R$_{GC}$ = 12~kpc) to 6 (R$_{GC}$ = 14~kpc)~\kmsperkpc without any inclination variations. However, the previously mentioned studies only indicate precession (or lack thereof) in the Galactic outer disk (R$_{GC}$ > 9~kpc), without any evidence of warping in the inner Galaxy.

Star clusters are essential in Galactic studies because they provide more precise distance and motion estimates compared to individual stellar tracers. However, only a few open clusters (OCs) located in the southern warp have been identified in previous study~\citep{cg20}. This is primarily due to the scarcity of distant OCs, making it difficult to explore the geometry of the Milky Way warp on a larger scale. Recently, we discovered nearly 1500 reliable open clusters through machine learning and visual inspection based on Gaia DR3~\citep{he23b}. This new result represents a substantial improvement over previous studies, which identified only one-forth of the present sample's distant OCs located more than 4~kpc away. This immense dataset presents an opportunity to explore the wider disk via OCs' study that were previously unattainable.
The primary objective of this study is to utilize this extensive sample of Gaia DR3-based OCs to gather vital geometric and kinematic information throughout the Galactic disk. By doing so, it may provide us with further insight into the Milky Way's warp.

Chapter~\ref{sec:data} introduces the selection process and the warp structure inferred from open clusters and classical Cepheids. 
Chapters~\ref{sec:geo} and~\ref{sec:kine} present the geometry and kinematic results, respectively. We analyze how orbital inclination, precession, and inclinational variation changes at different radii. The study presents an in-depth analysis of these findings, providing necessary insights into the warp structure traced by the sample.
Chapter~\ref{sec:discussion} summarizes the paper's main findings and presents potential future research opportunities. 
Finally, Appendix~\ref{sec:appendix_a} presents error analysis of the OC sample, and Appendix~\ref{sec:appendix_b} show a sketch map of elliptical inclined plane. 

\section{OC sample and warp structure}\label{sec:data}

Our study utilized a large sample of 2017 open clusters from ~\citet{cg20}, previously used to study the structure of the Milky Way disk within approximately 4~kpc of the Solar system in the Gaia DR2 era. We then cross-matched the member stars of these clusters in Gaia DR3, selecting only clusters with 20 or more member stars (a total of 1837 clusters). Additionally, we obtained new star clusters from Gaia DR2 and EDR3 with a sample size of over 500, including 615 in ~\citet{he21,he22a}, 628 in ~\citet{castro22}, and 1656 in ~\citet{he23a}. We cross-matched the member stars of these clusters in Gaia DR3 to obtain additional line-of-sight velocity information. We removed 138 duplicate clusters~\citep[from][]{he21,he22a,castro22} and 746 clusters with less than 20 member stars~\citep[mostly in][]{he23a}. In total, these clusters contain 3852 open clusters, among which 92$\%$ are within 4~kpc. Our study's sample provides a more extensive dataset for analyzing the structure of the Milky Way's disk.

\begin{figure*}
\begin{center}
\includegraphics[width=1.0\linewidth]{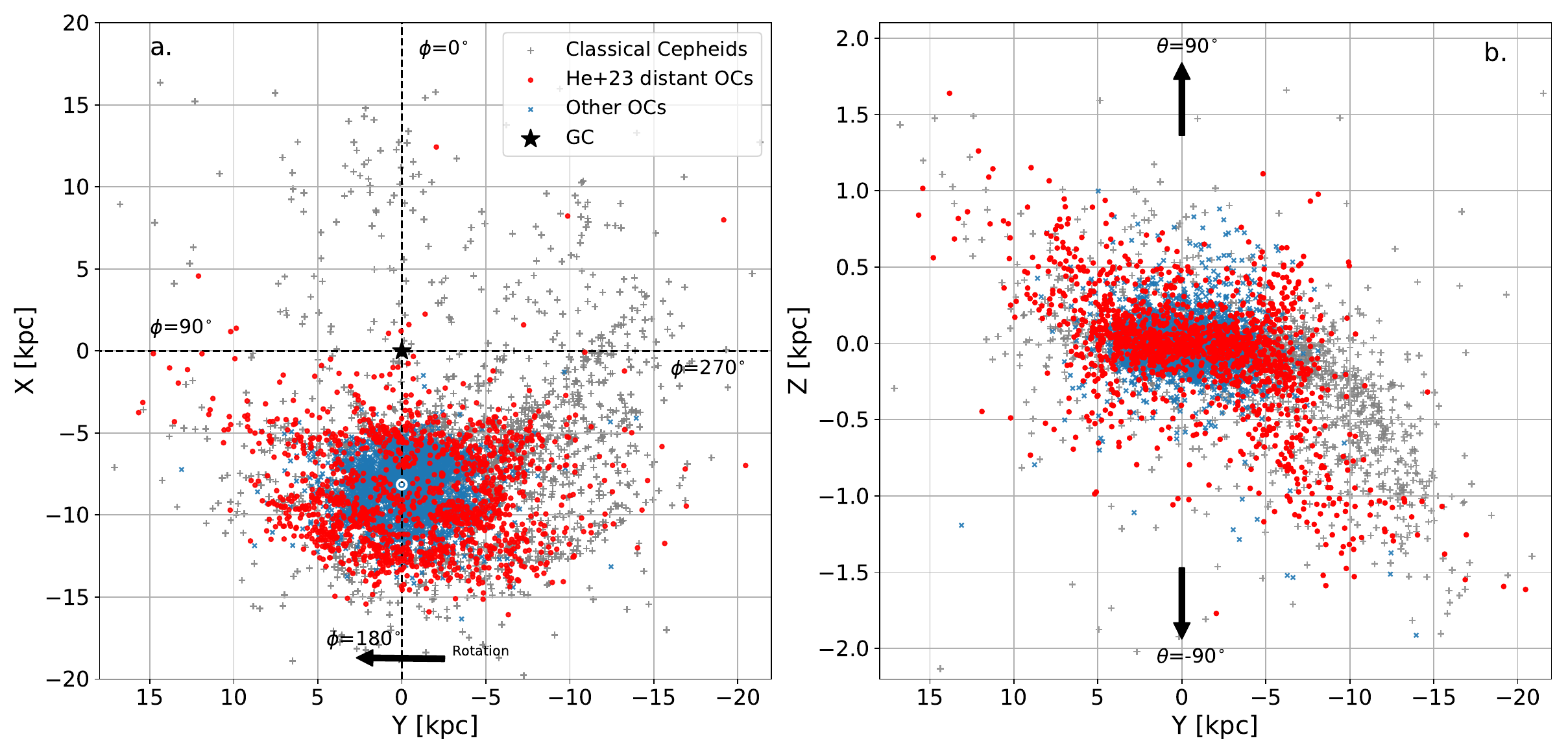}
\caption{
The distribution of open clusters in the XY plane is depicted in both the face-on (a.) and edge-on (b.) views, with Galactocentric coordinates ($\phi,\theta$) also displayed. In the left panel, the black arrow indicates the rotation direction of the Milky Way, and the position of the Solar system is (-8.15, 0) kpc, according to ~\citet{Reid19}. Red dots represent distant open clusters from~\citet{he23b}, while blue crosses denote other Gaia-based open clusters~\citep{cg20,castro22,he21,he22a,he23a}. For comparison, the gray pluses signify the CC sample ~\citep[][]{Skowron19}, showing a similar intuitive plot of the warp structure (based on CCs) from ~\citet[][]{Chen19,Lemasle22}; except for the southern warp at $\phi \sim~270^{\circ}$, which still lack of OC samples there.
}
\label{fig:xy_distribution}
\end{center}
\end{figure*}

Recently, a machine learning algorithm was employed by ~\citet{he23b} to identify and verify 2085 star clusters/candidates, out of which 1488 were reliable OCs (Type~1) verified through visual inspection. A total of 944 Type~1 OCs were located beyond 4~kpc, some with large extinction values (A$_V$~>~5~mag). Each star cluster underwent isochrone fitting and visual inspection, resulting in a more in-depth analysis of distant OCs' geometric structure.
The coordinates of the above nearby OC samples, combined with Type~1 distant OCs, were transformed into the Galactocentric coordinate system (more details in Appendix~\ref{sec:appendix_a}). A significant number of distant star clusters demonstrate the twisted/spiral structures of the Milky Way, as shown in Figure 1. This is the second celestial body type, after classical Cepheids (as shown in Figure~\ref{fig:xy_distribution}, grey symbols), to directly trace the twisted geometric structure. The highest scale height can reach over 1.5~kpc, both in the southern ($\theta$ < 0) and northern ($\theta$ > 0) Galaxy.
However, the edge-on image from the anti-Galactocentric line shown in Figure~\ref{fig:xy_distribution} only shows the warp in the outer disk. Are there any other tilted or warped structures in the inner regions?

%%%%%%%%%%%%%%%%%%%%%%%%%%figure%%%%%%%%%%%%%%%%%%%%%%%%%%%%%%%%%
%%%%%%%%%%%%%%%%%%%%%%%%%%figure%%%%%%%%%%%%%%%%%%%%%%%%%%%%%%%%%
%

\section{Disk inclination}\label{sec:geo}
\begin{figure*}
\begin{center}
\includegraphics[width=1.0\linewidth]{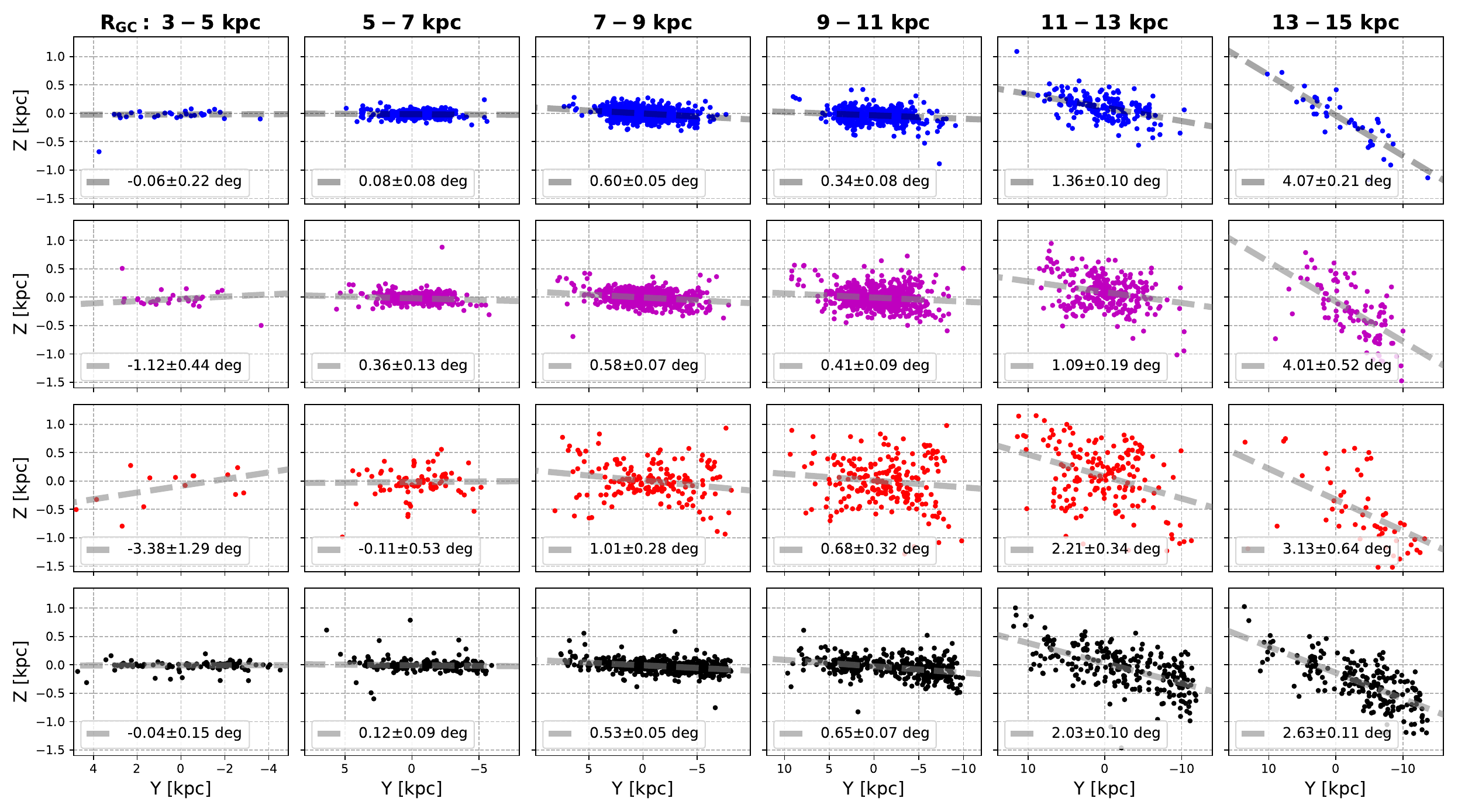}
\caption{
The edge-on views display Gaia-based open clusters (colored dots) and classical Cepheids (black dots) from ~\citet{Skowron19} located at varying Galactocentric radii around R$_{GC}$ = 4, 6, 8, 10, 12, and 14~kpc, all within a width range of $\pm$~1~kpc. The gray dashed line shows the linear regression of the inclination angle on the line of sight anti-Galactic center. Young (<~10$^8$~yr), middle-aged (10$^8$ - 10$^9$ yr), and old (>~10$^9$~yr) OCs are identified by the blue, magenta, and red symbols, respectively. As the radius increases, the thickness of the OC/CC disks gradually increase, displaying flaring characteristics. Additionally, compared to younger open clusters, the older ones tend to have higher scale heights, as stated in ~\citet{cg20}. It is essential to note that the LON are not situated at Y = 0 kpc, and the fitted $\theta_i$ is visible in Figure~\ref{fig:fit_vv}-a.
}
\label{fig:age_zz}
\end{center}
\end{figure*}

To examine the spatial distribution of open clusters inside and outside the Galactic disk, we present an edge-on view of the plane at different radii using a 2~kpc bin (Figure~\ref{fig:age_zz}), from the Galactic anticenter direction. The traced disc's degree of tilt by the cluster increases as the radius gets larger, demonstrating that the Milky Way's disc is tilted at various radii. Although the tilt is not very apparent at 6~kpc, a trend of tilt angle change can be observed.
Remarkably, the distribution of Cepheid variables exhibits the same tilt pattern, and their distances obtained through the period-luminosity relation indicate consistency with the geometric distribution derived from Gaia's parallax measurements. Our analysis suggests that the tilted disc structure is not unique to outer disk, as previously thought, but rather extends to other regions. Such findings could help to refine our understanding of the Milky Way's structure.

%%%%%%%%%%%%%%%%%%%%%%%%%%figure%%%%%%%%%%%%%%%%%%%%%%%%%%%%%%%%%

%%%%%%%%%%%%%%%%%%%%%%%%%%figure%%%%%%%%%%%%%%%%%%%%%%%%%%%%%%%%%
To shed more light on our findings, we categorized the OC samples into three sub-groups based on their age: young (YOCs < 100~Myr), middle-aged (MOCs: 0.1-1~Gyr), and old (OOCs > 1~Gyr) OCs. Interestingly, the inclination angles observed by different age tracers show only slight variations, indicating that YOCs and MOCs are highly consistent. On the other hand, OOCs usually show varying degrees of tilt; however, the trend of the tilt angle change is consistent with the other samples. We attribute this inconsistency among older clusters to radial migration or incomplete sub-samples.
As presented in Figure~\ref{fig:age_zz}, there is a consistent inclination angle of about 0.6 degrees on the orbit where the Sun is located. This value remains consistent for YOCs, MOCs, and CCs, with only a slightly larger inclination in OOCs. 
However, in the case of older clusters, the member stars in the main sequence exhibit a tendency to be fainter compared to their younger counterparts. This difference in brightness has the potential to result in notable variations in the observed distribution between the older OCs and the younger ones, particularly under heavy extinctions. Consequently, the apparent strong pattern observed in older samples, particularly within the 3 to 5 kpc range of R$_{GC}$ (Figure~\ref{fig:age_zz}), may not accurately reflect the true inclination.

\begin{figure*}
\begin{center}
\includegraphics[width=1.0\linewidth]{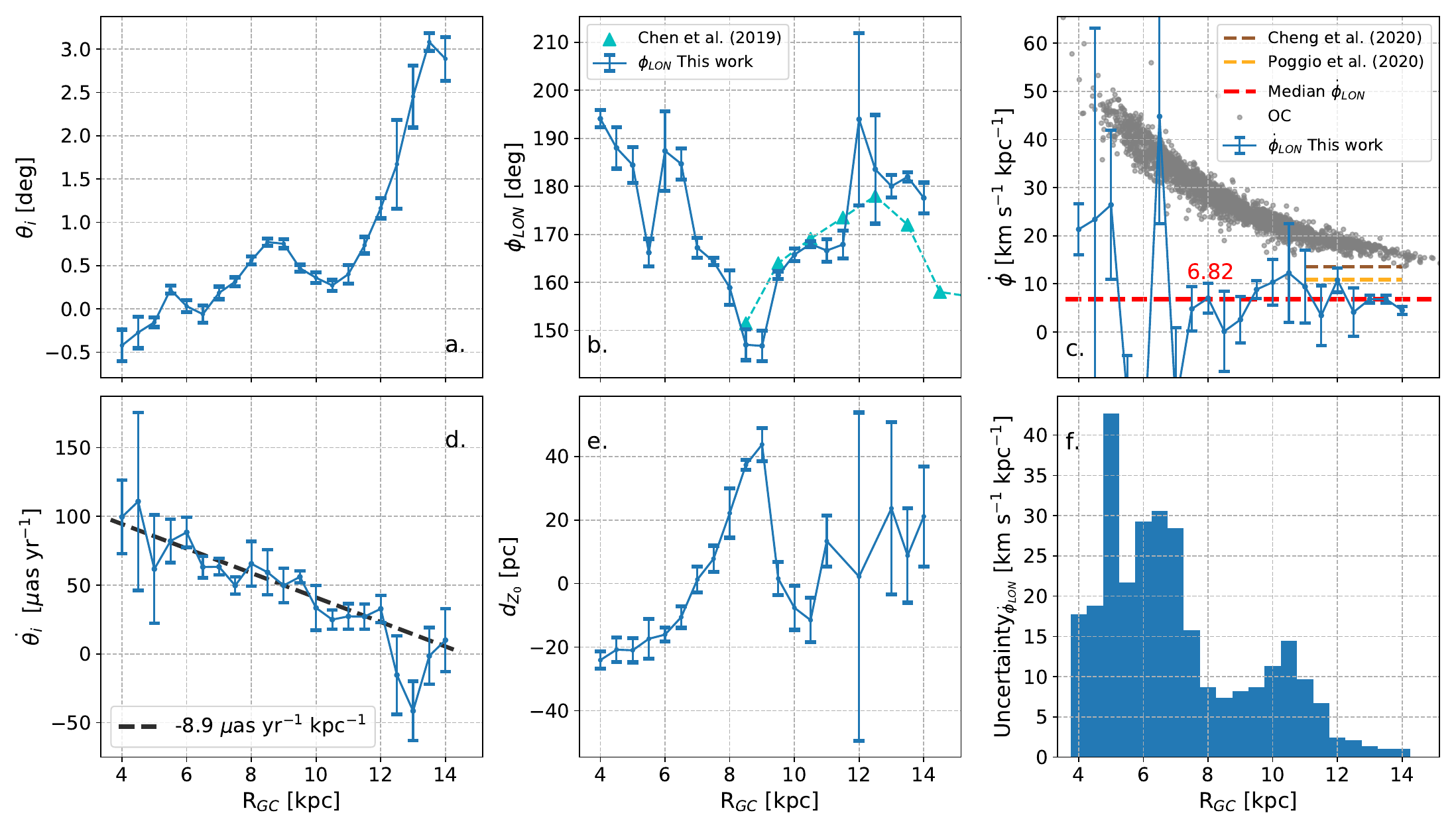}
\caption{
The derived parameters from Equation~\ref{eq1} and ~\ref{eq2} are presented with error bars accounting for uncertainties in OC sample age limitations. The parameters include: (a) inclination angle $\theta_i$, (b) LON positions $\phi_{LON}$, (c) precession rate $\dot{\phi}_{LON}$, (d) inclinational variation rate $\dot{\theta}_i$, (e) differential scale height d$_{z_0}$, and (f) uncertainty (from V$_\odot$) of $\dot{\phi}_{LON}$, plotted against Galactocentric radius R$_{GC}$. The typical uncertainty resulting from V$_\odot$ for $\dot{\theta}_i$ is 4.3~$\mu$as~yr$^{-1}$.
 }
\label{fig:fit_vv}
\end{center}
\end{figure*}

To calculate the Milky Way disc's inclination angle ($\theta_i$) and minimize uncertainties caused by bin selection and LON positions, we fitted OC sample values at different radii using a 0.5~kpc step and obtained the LON position from the OC kinematics (Section~\ref{sec:kine}). During this process, we discovered that the disk did not share a cohesive plane at different radii and that the scale height of the Solar system varied in different literature. Additionally, The lopsided warps were presented in various radius, particularly in the outer arm regions. We believe this is due to the objects' orbits not being circular but instead elliptical (discussed in Section~\ref{sec:discussion}). Nevertheless, to simplify our calculation, we used a circular orbit approximation and added $d_{Z_0}$ to eliminate orbital ellipticity/Z$_\odot$ impact on the scale height:
\begin{equation}\label{eq1}
Z (R_{GC}, \phi_{GC}, \phi_{LON})_{obs} = R_{GC} \times {\sin (\phi_{LON} - \phi_{GC} )} \times \sin \theta_i + d_{Z_0}
\end{equation}
Here, $\phi_{LON}$ and $\phi_{GC}$ represent the position of LON and OC samples in Galactocentric coordinates, and $Z_{obs}$ denotes the OCs' scale height.
Figure~\ref{fig:fit_vv}-a displays the inclination angle of the disk at different radii, and the error bars reflect the uncertainty in using OC samples of varying age limits. From the inner disk to the outer disk, the tilt angle gradually increases, with the absolute value of the tilt angle being approximately 0$^{\circ}$ near R$_{GC}$ = 6~kpc, and the tilt direction changes. The tilt angle experiences a slight decrease near R$_{GC}$ = 10~kpc, followed by a rapid increase to 3 degrees at R$_{GC}$ = 14~kpc. Unfortunately, the insufficient OC samples within R$_{GC}$ = 4~kpc leaves the orbital tilt in the Milky Way bulge unknown. Nevertheless, the limited information we do have suggests a negative trend in the inclination angle at that location.

\begin{figure*}
\begin{center}
\includegraphics[width=0.8\linewidth]{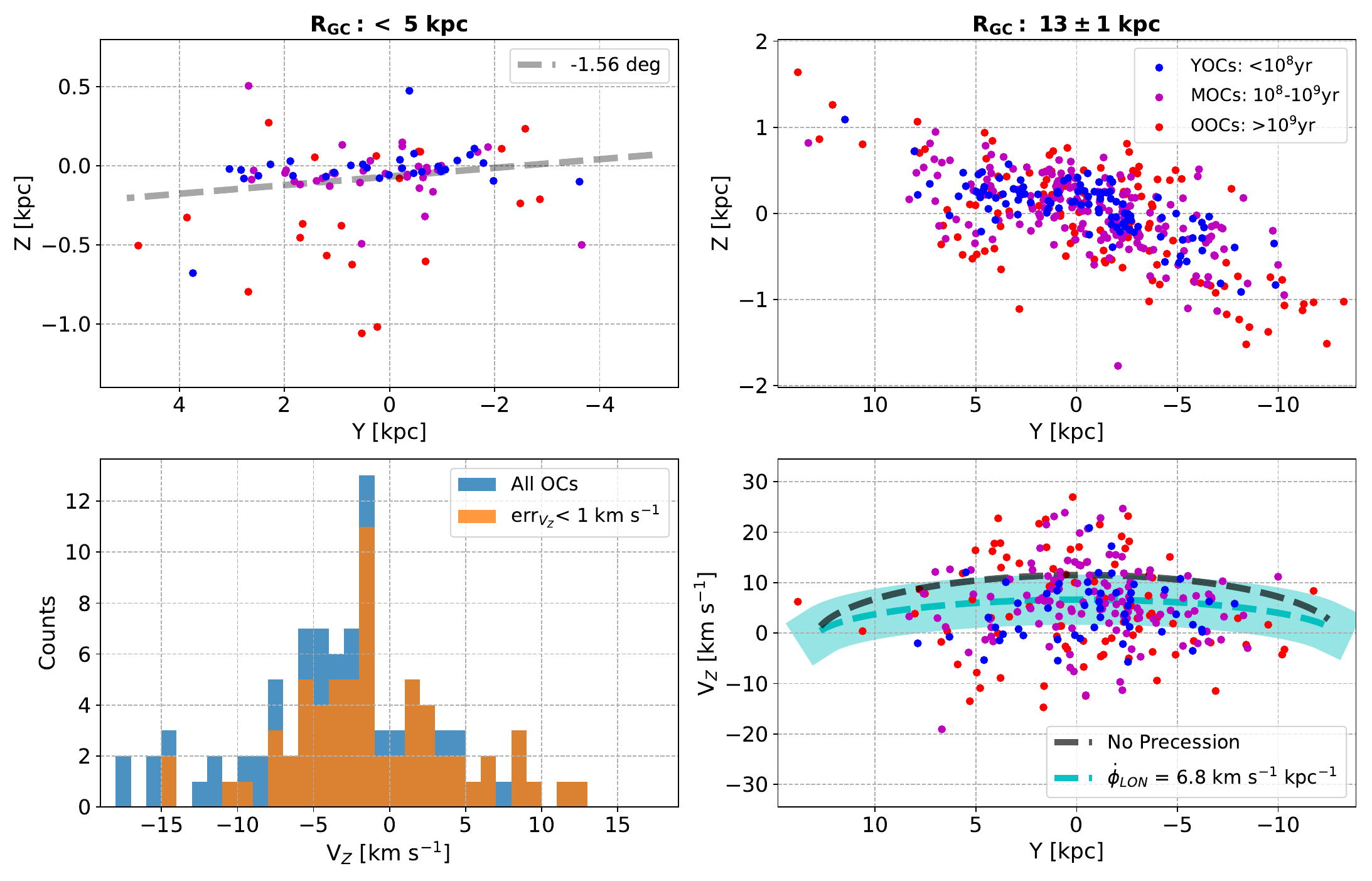}
\caption{
The edge-on views of OCs in the inner (left upper panel) and outer (right upper panel) Galactic disk, along with their corresponding vertical velocity distributions (displayed in the lower panels) are presented. A linear regression is applied (gray dashed line) to demonstrate the smaller inclination angles of the inner disk for clear visualization. Even though the inner disk does not exhibit the same significant tilt as the outer disk, a noticeable negative V$_Z$ is still observed, consistent with the direction of the tilt-induced velocity deviation. In the right lower panel, the black dashed line and cyan dashed line represent the vertical velocity when precession rate $\dot{\phi}_{LON}$ = 0 and 6.8 \kmsperkpc, respectively. It can be observed that the $\dot{\phi}_{LON}$ = 0 does not correspond to the observed V$_Z$ distribution, as supported by the study conducted by ~\citet{Poggio20} utilizing proper motions of giant stars.
}
\label{fig:yes_pre}
\end{center}
\end{figure*}

The consistent geometric features of tilt disks displayed by different sub-samples, as illustrated in Figure~\ref{fig:age_zz} and Figure~\ref{fig:fit_vv}, indicate that the samples utilized in this study are adequately representative for investigating the warp structures, especially for YOCs and MOCs. 
The degree of tilt at most radii appears to be less reliant on the sample selection. Additionally, as previously stated, orbital tilt is not solely reflected in geometric features; it also displays a vertical systemic velocity in kinematics, which is unaffected by extinction in the Galactic inner disk. As shown in Figure~\ref{fig:yes_pre}, a notable negative value is presented in the vertical velocity in the inner Galactic disc. This value is still maintained despite uncertainties regarding the Sun's peculiar motion, orbital precession, and inclinational variation in the Galaxy. When combined with their geometric features, these compelling pieces of evidence demonstrate that the orbit of the most inner regions of the Milky Way is also tilted, and the direction of this tilt is opposite to that of the outer disk.

%%%%%%%%%%%%%%%%%%%%%%%%%%figure%%%%%%%%%%%%%%%%%%%%%%%%%%%%%%%%%
%%%%%%%%%%%%%%%%%%%%%%%%%%figure%%%%%%%%%%%%%%%%%%%%%%%%%%%%%%%%%

%%%%%%%%%%%%%%%%%%%%%%%%%%figure%%%%%%%%%%%%%%%%%%%%%%%%%%%%%%%%%
%%%%%%%%%%%%%%%%%%%%%%%%%%figure%%%%%%%%%%%%%%%%%%%%%%%%%%%%%%%%%

\section{Galactic Motion}\label{sec:kine}

As mentioned above, a tilted disc can cause a vertical velocity component that reaches its maximum near the LON and gradually decrease towards zero at the highest point of the warp 90 degrees away from the northern/southern solstices. Moreover, a positive LON rotation (precession, $\dot{\phi}_{LON}$) can weaken the V$_{Z\odot}$ systematical velocity caused by this tilted orbit. Figure~\ref{fig:yes_pre} displays the measured results at R$_{GC}$ = 13~kpc, revealing that if only tilt and no precession existed, the observed vertical velocity would be higher. However, the observed velocity is consistent with a precession rate of $\dot{\phi}_{LON}$ = 6.8~\kmsperkpc, which is lower than what some previous studies~\citep{Poggio20,Cheng20} have reported. Nonetheless, this outcome is independent of any warping or kinematic model, confirming the presence of the precession of the Milky Way.

The Galactic coordinate system exhibits hints of rotation, ranging from 0.05 to 0.8~mas~yr$^{-1}$, based on various objects and observations~\citep[e.g.][]{Miyamoto98,Zhu00,Bobylev19}. ~\citet{Lindegren18} identified in Gaia data that the rotation is not more than 0.15~mas~yr$^{-1}$, showing 0.1~mas~yr$^{-1}$ when several VLBA sources were compared~\citep{Lindegren20}. Such a magnitude of frame rotation can also affect the proper motion $\mu_b$ or vertical velocity in Solar orbit systematically. For instance, a 50~$\mu$as~yr$^{-1}$ rate may cause 0 ($\phi_{LON}$ = 0$^{\circ}$) to $\sim\pm$~2~\kms ($\phi_{LON}$ =~$\mp$~90$^{\circ}$) vertical velocity around the Solar orbit. 
Additionally, it is reasonable to consider the variation of inclination since the inclination itself needs to increase/decrease to the current positions. Nevertheless, previous investigations on warping disks did not consider its effect, except for recent work~\citet{Dehnen23}. In this study, the precession $\dot{\phi}_{LON}$ and mean (since the orbit is not a rigid body) inclinational variation $\dot{\theta}_i$ are described by the simple equation:
\begin{equation}\label{eq2}
V_{Z} (R_{GC}, \phi_{GC}, \theta_i)_{obs} = \frac {|\theta_i|}{\theta_i} \times  (\frac {R_{GC} \times (\dot{\phi}_{GC} - \dot{\phi}_{LON})} {\sqrt{ \frac{1 + \tan^{-2}(\theta_i)} {\cos (\phi_{LON} - \phi_{GC} )} -1}} ) + \dot{\theta_i} \times R_{GC} \times \sin (\phi_{LON} - \phi_{GC}) \times \cos (\theta_i) - V_{Z \odot}
\end{equation}
%%%

The most significant uncertainty when calculating the precession rate is V$_{Z \odot}$ due to the 0.6-degree inclination near the Solar orbit, along with potential uncertainties from Solar peculiar motion, precession, and inclinational variation near the Sun. To compensate, an uncertainty of V$_{Z \odot}$ = 1~$\pm$~1~\kms was added. Using the OC kinematics, LON positions of all radii were fitted, leading to a wide distribution around $\phi$ = 170$^{\circ}$ (Figure~\ref{fig:fit_vv}-b). The LON positions in the outer Galactic disk ranges from $\phi$ =  145$^{\circ}$ (R$_{GC}$ = 8~kpc) to 180$^{\circ}$ (R$_{GC}$ = 14~kpc), consistent with the LON positions from CC samples~\citep{Chen19}, whereas the inner regions show an opposite trend. 
Additionally, our study revealed that the potential systematic movement associated with the rotation of the LON around the line-of-solstice is negligible, with an error bar of approximately ~$\sim$10~$\mu$as~yr$^{-1}$.

In the inner disk, despite a large sub-sample size, the small absolute value of the tilt angle (especially at R$_{GC}$ = 5 to 7~kpc, where the $\theta_i$ is close to zero) results in a significant influence of V$_{Z \odot}$ uncertainty on the outcome. A slight change in V$_{Z \odot}$ can cause a large differential $\dot{\phi}_{LON}$ (Figure~\ref{fig:fit_vv}-f). Nevertheless, considering the uncertainty, a precession rate greater than zero ( > 5~\kmsperkpc or greater) can be observed at R$_{GC}$ = 4~kpc, indicating the presence of precession characteristics in the inner disk, similar to the outer disk. Figure~\ref{fig:fit_vv}-d shows inclinational variation across the Milky Way disk, where $\dot{\theta}_i$ is higher in the inner disk and decreases with a slope of -8.9~$\mu$as~yr$^{-1}$~kpc$^{-1}$ towards the outer disk. The $\dot{\theta}_i$ in the region where the Solar system is located is 59.1~$\pm$~11.2$_{sample}$~$\pm$~7.7$_{V_{Z\odot}}$~$\mu$as~yr$^{-1}$, consistent with hints of previous findings~\citep{Lindegren18,Bobylev19,Lindegren20}. This inclinational variation across the disk suggests that the rotation of the LON reflects not only the motion of the Solar system but also the interior motion of the entire Galaxy.

\section{Discussion and Conclusion}\label{sec:discussion}

The findings presented in Figure~\ref{fig:fit_vv}-e reveal that the $d_{Z_0}$ values are not zero, suggesting that the median height of the tilted Milky Way disk is not on the same plane. This is likely an indication that the orbital motion of the disk in the Milky Way is a tilted ellipse plane instead of a circular orbit. Upon closer inspection, it was observed that $d_{Z_0}$ steadily increases from R$_{GC}$ = 4 to 9~kpc, which could be attributed to the varying position between the two foci at different orbital radii (as depicted in the Sketch map of Appendix~\ref{sec:appendix_b}). 
However, relying solely on ellipse orbit, the observed trend would suggest that eccentricity of the orbits in the inner disk are greater than 0.2 and could even be as high as 0.3, which is not significant enough to serve as evidence. High eccentricity may  cause the lopsided warp, but it also give rise to systematic radial motion near the solar orbit.
Therefore, it is necessary to conduct further research to gain a more profound understanding of this phenomenon.

This study presents observational evidence for precession and inclinational variation spanning both the inner and outer disks for the first time.  The analysis of different-aged OC samples produced significant results, revealing that the precession rates of the Galactic disk are lower than previously reported in the literature. This deviation is attributable to the fact that systematic vertical velocity, which affects the local standard of rest in Solar orbit, has been taken into account. Additionally, a global inclinational variation spread across the Galactic disk may suggest the disk galaxy is presently undergoing a shift in inclination. However, further observations and simulations are necessary to examine its origin better. It should be noted that the considerable rotation of the Solar orbit may impact high precision astrometry, necessitating additional analysis to identify any potential effects of precession and inclinational variation on the coordinate system.

%%%%%%
\section{Acknowledgements}
This work has made use of data from the European Space Agency (ESA) mission GAIA (\url{https://www.cosmos.esa.int/gaia}), processed by the GAIA Data Processing and Analysis Consortium (DPAC,\url{https://www.cosmos.esa.int/web/gaia/dpac/consortium}). Funding for the DPAC has been provided by national institutions, in particular the institutions participating in the GAIA Multilateral Agreement.
This work is supported by "Young Data Scientists" project of the National Astronomical Data Center, CAS (NADC2023YDS-07); and Fundamental Research Funds of China West Normal University (CWNU, No.21E030).

\appendix

\section{Data processing and error analysis}\label{sec:appendix_a}
We employed the Gaia DR3 data to derive the cluster  position and velocity components from the member stars of 5340 open clusters. To do so, we first calculated the weighted means of the member stars' parameters and then adjusted the parallax by adding 0.017 mas~\citep[parallax zero point,][]{Lindegren21,Fabricius21}. By considering the Solar system's position: [R$_{GC\odot}$, Z$_\odot$] are [8.15~kpc, 5.5~pc], and Solar peculiar motion: [U$_{\odot}$, V$_{\odot}$, W$_{\odot}$] are [10.6, 10.7, 7.6]~\kms, and V$_{\phi0}$ = 236~\kms  ~\citep{Reid19}, we transformed these values into a coordinate system centered on the Galactic center (R$_{GC}$, $\phi$, Z) and a three-dimensional velocity (V$_{R_{GC}}$,  V$_\phi$, V$_Z$). It is worth noting that 1269 of the 5340 clusters did not have radial velocities, and we estimated the median value by selecting OCs in ($l \pm$~10$^{\circ}$, $b \pm$~5$^{\circ}$, $\varpi \pm$~0.05~mas) and set the error to maximum of 50~\kms and velocity dispersion of the sample.

\begin{figure*}
\begin{center}
\includegraphics[width=0.8\linewidth]{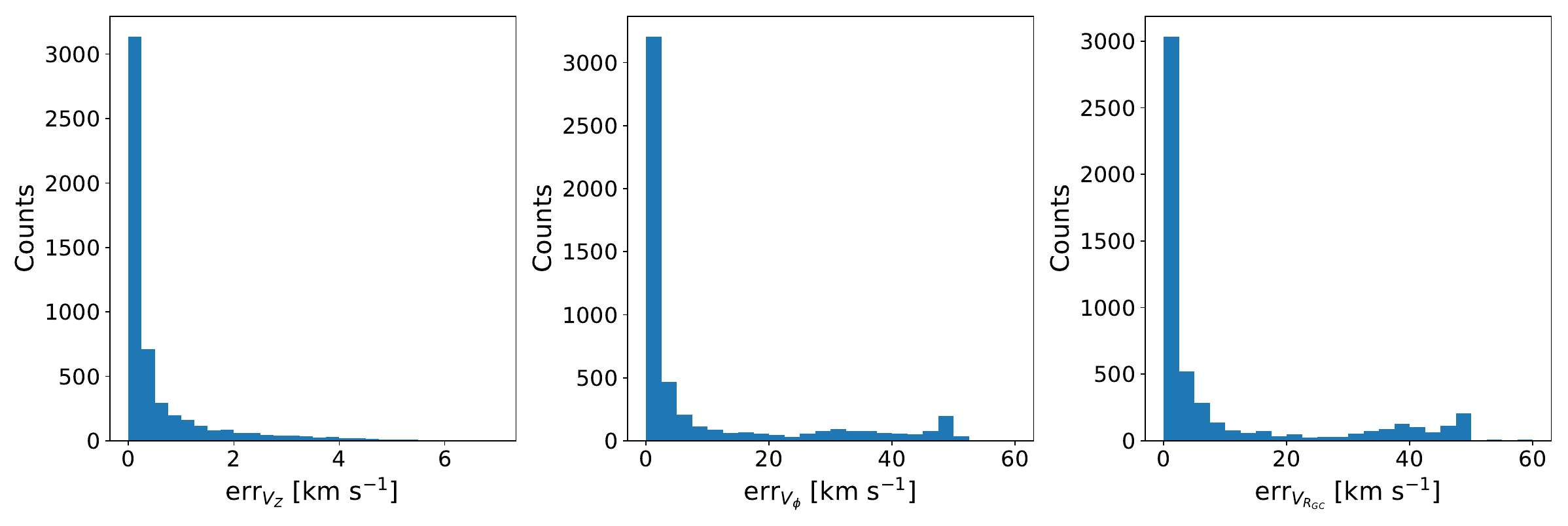}
\caption{Statistical distribution of velocity component errors presented in the Galactocentric coordinate system,  using 1~\kms bin size for V$_Z$ and 2.5~\kms bin size for V$_\phi$ and V$_{R_{GC}}$.}
\label{fig:err}
\end{center}
\end{figure*}

It is worth mentioning that the uncertainty of the vertical velocity component V$_\odot$ had the most substantial impact on our results. Given the variation in W$_\odot$ values across different studies~\citep[e.g., 7.6, 8.6, and 6.9~\kms from VLBA observations, Gaia data, and others, respectively.][]{Reid19,gc20210,Semczuk23}, we included the effect of this uncertainty in our calculation (Section~\ref{sec:kine}). Furthermore, we added $d_{Z_0}$ to reduce the impact of the inclinational ecliptic orbit and scale height uncertainty in the Solar system (Section~\ref{sec:geo}).
To fit for parameters in Figure~\ref{fig:fit_vv}, we selected an open cluster sample with err$_{V_Z} $< 1~\kms and err$_{V_\phi }$ < 2.5~\kms, which is equivalent to the typical intrinsic velocity dispersion of OCs~\citep[around 1~\kms or less,][]{Mermilliod09}, without placing any distance-based constraints. However, we accounted for distance errors while estimating velocity errors, resulting in the exclusion of some star clusters with large distance errors in the velocity selections. Furthermore, we employed a bin size of 2~kpc while fitting the data, effectively minimizing the effects of distance errors (distance error <~500~pc for 93$\%$ of the clusters). Figure~\ref{fig:err} depicts the histograms of the derived velocity errors.

%%%%%%%%%%%%%%%%%%%%%%%%%%figure%%%%%%%%%%%%%%%%%%%%%%%%%%%%%%%%%
%%%%%%%%%%%%%%%%%%%%%%%%%%figure%%%%%%%%%%%%%%%%%%%%%%%%%%%%%%%%%

\section{Sketch map of elliptical inclined galactic plane}\label{sec:appendix_b}
\begin{figure*}
\begin{center}
\includegraphics[width=0.8\linewidth]{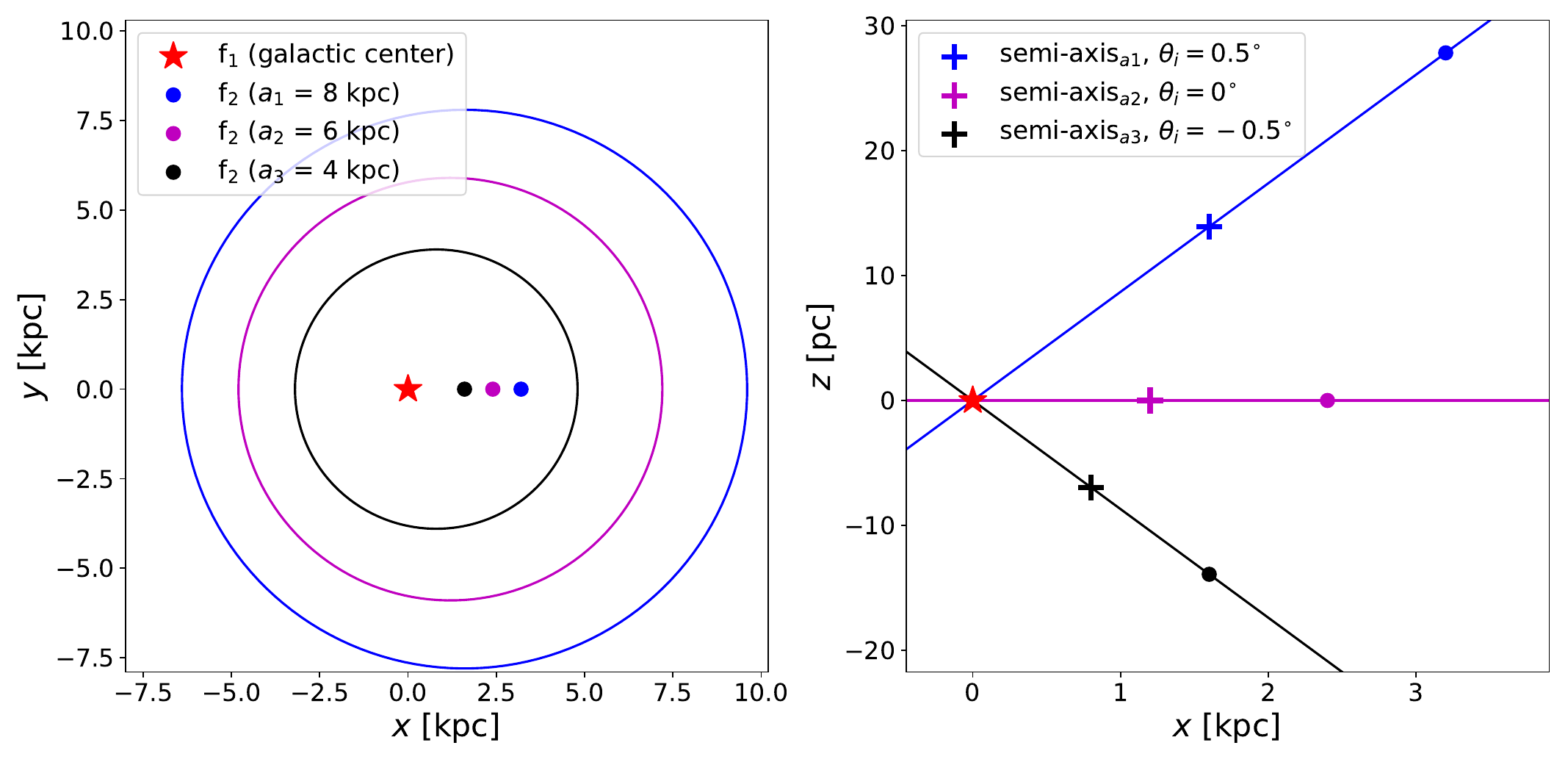}
\caption{
Sketch map of scale height in the elliptical inclined galactic plane. The left panel displays the elliptical plane at different semi-major axis, represented by the function: $x^2/a^2+y^2/b^2=1$, where $f_1$ (galactic center) and $f_2$ are the foci, and the eccentricity is set to 0.2. Right panel demonstrates the differential scale height of the $f_2$ and semi-minor axis, with the inclination angles ($\theta_i$) set at 0.5 (blue), 0 (magenta), and -0.5 (black) degrees.
}
\label{fig:f1f2}
\end{center}
\end{figure*}

Figure~\ref{fig:f1f2} depicts the sketch map of the scale height in the elliptical inclined galactic plane. The corresponding median scale height follows the trend of $d_{Z_0}$ in R$_{GC}$ = 4 to 11~kpc, as seen in Figure~\ref{fig:fit_vv}-e.

%%%%%%%%%%%%%%%%%%%%%%%%%%figure%%%%%%%%%%%%%%%%%%%%%%%%%%%%%%%%%
%%%%%%%%%%%%%%%%%%%%%%%%%%figure%%%%%%%%%%%%%%%%%%%%%%%%%%%%%%%%%

\bibliographystyle{aasjournal} 
\bibliography{mw} 

\begin{thebibliography}{}
\expandafter\ifx\csname natexlab\endcsname\relax\def\natexlab#1{#1}\fi

\bibitem[{{Bobylev} \& {Bajkova}(2019)}]{Bobylev19}
{Bobylev}, V.~V., \& {Bajkova}, A.~T. 2019, Astronomy Letters, 45, 208

\bibitem[{{Bosma}(1981)}]{Bosma81}
{Bosma}, A. 1981, \aj, 86, 1791

\bibitem[{{Burke}(1957)}]{Burke57}
{Burke}, B.~F. 1957, \aj, 62, 90

\bibitem[{{Burton}(1988)}]{Burton88}
{Burton}, W.~B. 1988, in Galactic and Extragalactic Radio Astronomy, ed. K.~I.
  {Kellermann} \& G.~L. {Verschuur}, 295--358

\bibitem[{{Cantat-Gaudin} {et~al.}(2020){Cantat-Gaudin}, {Anders},
  {Castro-Ginard}, {Jordi}, {Romero-G{\'o}mez}, {Soubiran}, {Casamiquela},
  {Tarricq}, {Moitinho}, {Vallenari}, {Bragaglia}, {Krone-Martins}, \&
  {Kounkel}}]{cg20}
{Cantat-Gaudin}, T., {Anders}, F., {Castro-Ginard}, A., {et~al.} 2020, \aap,
  640, A1

\bibitem[{{Castro-Ginard} {et~al.}(2022){Castro-Ginard}, {Jordi}, {Luri},
  {Cantat-Gaudin}, {Carrasco}, {Casamiquela}, {Anders},
  {Balaguer-N{\'u}{\~n}ez}, \& {Badia}}]{castro22}
{Castro-Ginard}, A., {Jordi}, C., {Luri}, X., {et~al.} 2022, \aap, 661, A118

\bibitem[{{Chen} {et~al.}(2019){Chen}, {Wang}, {Deng}, {de Grijs}, {Liu}, \&
  {Tian}}]{Chen19}
{Chen}, X., {Wang}, S., {Deng}, L., {et~al.} 2019, Nature Astronomy, 3, 320

\bibitem[{{Cheng} {et~al.}(2020){Cheng}, {Anguiano}, {Majewski}, {Hayes},
  {Arras}, {Chiappini}, {Hasselquist}, {de Andrade Queiroz}, {Nitschelm},
  {Garc{\'\i}a-Hern{\'a}ndez}, {Lane}, {Roman-Lopes}, \&
  {Frinchaboy}}]{Cheng20}
{Cheng}, X., {Anguiano}, B., {Majewski}, S.~R., {et~al.} 2020, \apj, 905, 49

\bibitem[{{Chrob{\'a}kov{\'a}} \& {L{\'o}pez-Corredoira}(2021)}]{Corredoira21}
{Chrob{\'a}kov{\'a}}, {\v{Z}}., \& {L{\'o}pez-Corredoira}, M. 2021, \apj, 912,
  130

\bibitem[{{Dame} \& {Thaddeus}(2011)}]{Dame11}
{Dame}, T.~M., \& {Thaddeus}, P. 2011, \apjl, 734, L24

\bibitem[{{Dehnen} {et~al.}(2023){Dehnen}, {Semczuk}, \&
  {Sch{\"o}nrich}}]{Dehnen23}
{Dehnen}, W., {Semczuk}, M., \& {Sch{\"o}nrich}, R. 2023, \mnras,
  arXiv:2305.09343

\bibitem[{{Fabricius} {et~al.}(2021){Fabricius}, {Luri}, {Arenou}, {Babusiaux},
  {Helmi}, {Muraveva}, {Reyl{\'e}}, {Spoto}, {Vallenari}, {Antoja}, {Balbinot},
  {Barache}, {Bauchet}, {Bragaglia}, {Busonero}, {Cantat-Gaudin}, {Carrasco},
  {Diakit{\'e}}, {Fabrizio}, {Figueras}, {Garcia-Gutierrez}, {Garofalo},
  {Jordi}, {Kervella}, {Khanna}, {Leclerc}, {Licata}, {Lambert}, {Marrese},
  {Masip}, {Ramos}, {Robichon}, {Robin}, {Romero-G{\'o}mez}, {Rubele}, \&
  {Weiler}}]{Fabricius21}
{Fabricius}, C., {Luri}, X., {Arenou}, F., {et~al.} 2021, \aap, 649, A5

\bibitem[{{Gaia Collaboration} {et~al.}(2018){Gaia Collaboration}, {Brown},
  {Vallenari}, {Prusti}, {de Bruijne}, {Babusiaux}, {Bailer-Jones}, {Biermann},
  {Evans}, {Eyer}, {Jansen}, {Jordi}, {Klioner}, {Lammers}, {Lindegren},
  {Luri}, {Mignard}, {Panem}, {Pourbaix}, {Randich}, {Sartoretti}, {Siddiqui},
  {Soubiran}, {van Leeuwen}, {Walton}, {Arenou}, {Bastian}, {Cropper},
  {Drimmel}, {Katz}, {Lattanzi}, {Bakker}, {Cacciari}, {Casta{\~n}eda},
  {Chaoul}, {Cheek}, {De Angeli}, {Fabricius}, {Guerra}, {Holl}, {Masana},
  {Messineo}, {Mowlavi}, {Nienartowicz}, {Panuzzo}, {Portell}, {Riello},
  {Seabroke}, {Tanga}, {Th{\'e}venin}, {Gracia-Abril}, {Comoretto},
  {Garcia-Reinaldos}, {Teyssier}, {Altmann}, {Andrae}, {Audard},
  {Bellas-Velidis}, {Benson}, {Berthier}, {Blomme}, {Burgess}, {Busso},
  {Carry}, {Cellino}, {Clementini}, {Clotet}, {Creevey}, {Davidson}, {De
  Ridder}, {Delchambre}, {Dell'Oro}, {Ducourant},
  {Fern{\'a}ndez-Hern{\'a}ndez}, {Fouesneau}, {Fr{\'e}mat}, {Galluccio},
  {Garc{\'\i}a-Torres}, {Gonz{\'a}lez-N{\'u}{\~n}ez}, {Gonz{\'a}lez-Vidal},
  {Gosset}, {Guy}, {Halbwachs}, {Hambly}, {Harrison}, {Hern{\'a}ndez},
  {Hestroffer}, {Hodgkin}, {Hutton}, {Jasniewicz}, {Jean-Antoine-Piccolo},
  {Jordan}, {Korn}, {Krone-Martins}, {Lanzafame}, {Lebzelter}, {L{\"o}ffler},
  {Manteiga}, {Marrese}, {Mart{\'\i}n-Fleitas}, {Moitinho}, {Mora}, {Muinonen},
  {Osinde}, {Pancino}, {Pauwels}, {Petit}, {Recio-Blanco}, {Richards},
  {Rimoldini}, {Robin}, {Sarro}, {Siopis}, {Smith}, {Sozzetti}, {S{\"u}veges},
  {Torra}, {van Reeven}, {Abbas}, {Abreu Aramburu}, {Accart}, {Aerts},
  {Altavilla}, {{\'A}lvarez}, {Alvarez}, {Alves}, {Anderson}, {Andrei},
  {Anglada Varela}, {Antiche}, {Antoja}, {Arcay}, {Astraatmadja}, {Bach},
  {Baker}, {Balaguer-N{\'u}{\~n}ez}, {Balm}, {Barache}, {Barata}, {Barbato},
  {Barblan}, {Barklem}, {Barrado}, {Barros}, {Barstow}, {Bartholom{\'e}
  Mu{\~n}oz}, {Bassilana}, {Becciani}, {Bellazzini}, {Berihuete}, {Bertone},
  {Bianchi}, {Bienaym{\'e}}, {Blanco-Cuaresma}, {Boch}, {Boeche}, {Bombrun},
  {Borrachero}, {Bossini}, {Bouquillon}, {Bourda}, {Bragaglia}, {Bramante},
  {Breddels}, {Bressan}, {Brouillet}, {Br{\"u}semeister}, {Brugaletta},
  {Bucciarelli}, {Burlacu}, {Busonero}, {Butkevich}, {Buzzi}, {Caffau},
  {Cancelliere}, {Cannizzaro}, {Cantat-Gaudin}, {Carballo}, {Carlucci},
  {Carrasco}, {Casamiquela}, {Castellani}, {Castro-Ginard}, {Charlot},
  {Chemin}, {Chiavassa}, {Cocozza}, {Costigan}, {Cowell}, {Crifo}, {Crosta},
  {Crowley}, {Cuypers}, {Dafonte}, {Damerdji}, {Dapergolas}, {David}, {David},
  {de Laverny}, {De Luise}, {De March}, {de Martino}, {de Souza}, {de Torres},
  {Debosscher}, {del Pozo}, {Delbo}, {Delgado}, {Delgado}, {Di Matteo},
  {Diakite}, {Diener}, {Distefano}, {Dolding}, {Drazinos}, {Dur{\'a}n},
  {Edvardsson}, {Enke}, {Eriksson}, {Esquej}, {Eynard Bontemps}, {Fabre},
  {Fabrizio}, {Faigler}, {Falc{\~a}o}, {Farr{\`a}s Casas}, {Federici},
  {Fedorets}, {Fernique}, {Figueras}, {Filippi}, {Findeisen}, {Fonti},
  {Fraile}, {Fraser}, {Fr{\'e}zouls}, {Gai}, {Galleti}, {Garabato},
  {Garc{\'\i}a-Sedano}, {Garofalo}, {Garralda}, {Gavel}, {Gavras}, {Gerssen},
  {Geyer}, {Giacobbe}, {Gilmore}, {Girona}, {Giuffrida}, {Glass}, {Gomes},
  {Granvik}, {Gueguen}, {Guerrier}, {Guiraud}, {Guti{\'e}rrez-S{\'a}nchez},
  {Haigron}, {Hatzidimitriou}, {Hauser}, {Haywood}, {Heiter}, {Helmi}, {Heu},
  {Hilger}, {Hobbs}, {Hofmann}, {Holland}, {Huckle}, {Hypki}, {Icardi},
  {Jan{\ss}en}, {Jevardat de Fombelle}, {Jonker}, {Juh{\'a}sz}, {Julbe},
  {Karampelas}, {Kewley}, {Klar}, {Kochoska}, {Kohley}, {Kolenberg},
  {Kontizas}, {Kontizas}, {Koposov}, {Kordopatis}, {Kostrzewa-Rutkowska},
  {Koubsky}, {Lambert}, {Lanza}, {Lasne}, {Lavigne}, {Le Fustec}, {Le
  Poncin-Lafitte}, {Lebreton}, {Leccia}, {Leclerc}, {Lecoeur-Taibi},
  {Lenhardt}, {Leroux}, {Liao}, {Licata}, {Lindstr{\o}m}, {Lister}, {Livanou},
  {Lobel}, {L{\'o}pez}, {Managau}, {Mann}, {Mantelet}, {Marchal}, {Marchant},
  {Marconi}, {Marinoni}, {Marschalk{\'o}}, {Marshall}, {Martino}, {Marton},
  {Mary}, {Massari}, {Matijevi{\v{c}}}, {Mazeh}, {McMillan}, {Messina},
  {Michalik}, {Millar}, {Molina}, {Molinaro}, {Moln{\'a}r}, {Montegriffo},
  {Mor}, {Morbidelli}, {Morel}, {Morris}, {Mulone}, {Muraveva}, {Musella},
  {Nelemans}, {Nicastro}, {Noval}, {O'Mullane}, {Ord{\'e}novic},
  {Ord{\'o}{\~n}ez-Blanco}, {Osborne}, {Pagani}, {Pagano}, {Pailler},
  {Palacin}, {Palaversa}, {Panahi}, {Pawlak}, {Piersimoni}, {Pineau}, {Plachy},
  {Plum}, {Poggio}, {Poujoulet}, {Pr{\v{s}}a}, {Pulone}, {Racero}, {Ragaini},
  {Rambaux}, {Ramos-Lerate}, {Regibo}, {Reyl{\'e}}, {Riclet}, {Ripepi}, {Riva},
  {Rivard}, {Rixon}, {Roegiers}, {Roelens}, {Romero-G{\'o}mez}, {Rowell},
  {Royer}, {Ruiz-Dern}, {Sadowski}, {Sagrist{\`a} Sell{\'e}s}, {Sahlmann},
  {Salgado}, {Salguero}, {Sanna}, {Santana-Ros}, {Sarasso}, {Savietto},
  {Schultheis}, {Sciacca}, {Segol}, {Segovia}, {S{\'e}gransan}, {Shih},
  {Siltala}, {Silva}, {Smart}, {Smith}, {Solano}, {Solitro}, {Sordo}, {Soria
  Nieto}, {Souchay}, {Spagna}, {Spoto}, {Stampa}, {Steele},
  {Steidelm{\"u}ller}, {Stephenson}, {Stoev}, {Suess}, {Surdej}, {Szabados},
  {Szegedi-Elek}, {Tapiador}, {Taris}, {Tauran}, {Taylor}, {Teixeira},
  {Terrett}, {Teyssand ier}, {Thuillot}, {Titarenko}, {Torra Clotet}, {Turon},
  {Ulla}, {Utrilla}, {Uzzi}, {Vaillant}, {Valentini}, {Valette}, {van Elteren},
  {Van Hemelryck}, {van Leeuwen}, {Vaschetto}, {Vecchiato}, {Veljanoski},
  {Viala}, {Vicente}, {Vogt}, {von Essen}, {Voss}, {Votruba}, {Voutsinas},
  {Walmsley}, {Weiler}, {Wertz}, {Wevers}, {Wyrzykowski}, {Yoldas},
  {{\v{Z}}erjal}, {Ziaeepour}, {Zorec}, {Zschocke}, {Zucker}, {Zurbach}, \&
  {Zwitter}}]{gaiadr2}
{Gaia Collaboration}, {Brown}, A.~G.~A., {Vallenari}, A., {et~al.} 2018, \aap,
  616, A1

\bibitem[{{Gaia Collaboration} {et~al.}(2021){Gaia Collaboration}, {Brown},
  {Vallenari}, {Prusti}, {de Bruijne}, {Babusiaux}, {Biermann}, {Creevey},
  {Evans}, {Eyer}, {Hutton}, {Jansen}, {Jordi}, {Klioner}, {Lammers},
  {Lindegren}, {Luri}, {Mignard}, {Panem}, {Pourbaix}, {Randich}, {Sartoretti},
  {Soubiran}, {Walton}, {Arenou}, {Bailer-Jones}, {Bastian}, {Cropper},
  {Drimmel}, {Katz}, {Lattanzi}, {van Leeuwen}, {Bakker}, {Cacciari},
  {Casta{\~n}eda}, {De Angeli}, {Ducourant}, {Fabricius}, {Fouesneau},
  {Fr{\'e}mat}, {Guerra}, {Guerrier}, {Guiraud}, {Jean-Antoine Piccolo},
  {Masana}, {Messineo}, {Mowlavi}, {Nicolas}, {Nienartowicz}, {Pailler},
  {Panuzzo}, {Riclet}, {Roux}, {Seabroke}, {Sordo}, {Tanga}, {Th{\'e}venin},
  {Gracia-Abril}, {Portell}, {Teyssier}, {Altmann}, {Andrae}, {Bellas-Velidis},
  {Benson}, {Berthier}, {Blomme}, {Brugaletta}, {Burgess}, {Busso}, {Carry},
  {Cellino}, {Cheek}, {Clementini}, {Damerdji}, {Davidson}, {Delchambre},
  {Dell'Oro}, {Fern{\'a}ndez-Hern{\'a}ndez}, {Galluccio}, {Garc{\'\i}a-Lario},
  {Garcia-Reinaldos}, {Gonz{\'a}lez-N{\'u}{\~n}ez}, {Gosset}, {Haigron},
  {Halbwachs}, {Hambly}, {Harrison}, {Hatzidimitriou}, {Heiter},
  {Hern{\'a}ndez}, {Hestroffer}, {Hodgkin}, {Holl}, {Jan{\ss}en}, {Jevardat de
  Fombelle}, {Jordan}, {Krone-Martins}, {Lanzafame}, {L{\"o}ffler}, {Lorca},
  {Manteiga}, {Marchal}, {Marrese}, {Moitinho}, {Mora}, {Muinonen}, {Osborne},
  {Pancino}, {Pauwels}, {Petit}, {Recio-Blanco}, {Richards}, {Riello},
  {Rimoldini}, {Robin}, {Roegiers}, {Rybizki}, {Sarro}, {Siopis}, {Smith},
  {Sozzetti}, {Ulla}, {Utrilla}, {van Leeuwen}, {van Reeven}, {Abbas}, {Abreu
  Aramburu}, {Accart}, {Aerts}, {Aguado}, {Ajaj}, {Altavilla}, {{\'A}lvarez},
  {{\'A}lvarez Cid-Fuentes}, {Alves}, {Anderson}, {Anglada Varela}, {Antoja},
  {Audard}, {Baines}, {Baker}, {Balaguer-N{\'u}{\~n}ez}, {Balbinot}, {Balog},
  {Barache}, {Barbato}, {Barros}, {Barstow}, {Bartolom{\'e}}, {Bassilana},
  {Bauchet}, {Baudesson-Stella}, {Becciani}, {Bellazzini}, {Bernet}, {Bertone},
  {Bianchi}, {Blanco-Cuaresma}, {Boch}, {Bombrun}, {Bossini}, {Bouquillon},
  {Bragaglia}, {Bramante}, {Breedt}, {Bressan}, {Brouillet}, {Bucciarelli},
  {Burlacu}, {Busonero}, {Butkevich}, {Buzzi}, {Caffau}, {Cancelliere},
  {C{\'a}novas}, {Cantat-Gaudin}, {Carballo}, {Carlucci}, {Carnerero},
  {Carrasco}, {Casamiquela}, {Castellani}, {Castro-Ginard}, {Castro Sampol},
  {Chaoul}, {Charlot}, {Chemin}, {Chiavassa}, {Cioni}, {Comoretto}, {Cooper},
  {Cornez}, {Cowell}, {Crifo}, {Crosta}, {Crowley}, {Dafonte}, {Dapergolas},
  {David}, {David}, {de Laverny}, {De Luise}, {De March}, {De Ridder}, {de
  Souza}, {de Teodoro}, {de Torres}, {del Peloso}, {del Pozo}, {Delbo},
  {Delgado}, {Delgado}, {Delisle}, {Di Matteo}, {Diakite}, {Diener},
  {Distefano}, {Dolding}, {Eappachen}, {Edvardsson}, {Enke}, {Esquej}, {Fabre},
  {Fabrizio}, {Faigler}, {Fedorets}, {Fernique}, {Fienga}, {Figueras},
  {Fouron}, {Fragkoudi}, {Fraile}, {Franke}, {Gai}, {Garabato},
  {Garcia-Gutierrez}, {Garc{\'\i}a-Torres}, {Garofalo}, {Gavras}, {Gerlach},
  {Geyer}, {Giacobbe}, {Gilmore}, {Girona}, {Giuffrida}, {Gomel}, {Gomez},
  {Gonzalez-Santamaria}, {Gonz{\'a}lez-Vidal}, {Granvik},
  {Guti{\'e}rrez-S{\'a}nchez}, {Guy}, {Hauser}, {Haywood}, {Helmi}, {Hidalgo},
  {Hilger}, {H{\l}adczuk}, {Hobbs}, {Holland}, {Huckle}, {Jasniewicz},
  {Jonker}, {Juaristi Campillo}, {Julbe}, {Karbevska}, {Kervella}, {Khanna},
  {Kochoska}, {Kontizas}, {Kordopatis}, {Korn}, {Kostrzewa-Rutkowska},
  {Kruszy{\'n}ska}, {Lambert}, {Lanza}, {Lasne}, {Le Campion}, {Le Fustec},
  {Lebreton}, {Lebzelter}, {Leccia}, {Leclerc}, {Lecoeur-Taibi}, {Liao},
  {Licata}, {Lindstr{\o}m}, {Lister}, {Livanou}, {Lobel}, {Madrero Pardo},
  {Managau}, {Mann}, {Marchant}, {Marconi}, {Marcos Santos}, {Marinoni},
  {Marocco}, {Marshall}, {Martin Polo}, {Mart{\'\i}n-Fleitas}, {Masip},
  {Massari}, {Mastrobuono-Battisti}, {Mazeh}, {McMillan}, {Messina},
  {Michalik}, {Millar}, {Mints}, {Molina}, {Molinaro}, {Moln{\'a}r},
  {Montegriffo}, {Mor}, {Morbidelli}, {Morel}, {Morris}, {Mulone}, {Munoz},
  {Muraveva}, {Murphy}, {Musella}, {Noval}, {Ord{\'e}novic}, {Orr{\`u}},
  {Osinde}, {Pagani}, {Pagano}, {Palaversa}, {Palicio}, {Panahi}, {Pawlak},
  {Pe{\~n}alosa Esteller}, {Penttil{\"a}}, {Piersimoni}, {Pineau}, {Plachy},
  {Plum}, {Poggio}, {Poretti}, {Poujoulet}, {Pr{\v{s}}a}, {Pulone}, {Racero},
  {Ragaini}, {Rainer}, {Raiteri}, {Rambaux}, {Ramos}, {Ramos-Lerate}, {Re
  Fiorentin}, {Regibo}, {Reyl{\'e}}, {Ripepi}, {Riva}, {Rixon}, {Robichon},
  {Robin}, {Roelens}, {Rohrbasser}, {Romero-G{\'o}mez}, {Rowell}, {Royer},
  {Rybicki}, {Sadowski}, {Sagrist{\`a} Sell{\'e}s}, {Sahlmann}, {Salgado},
  {Salguero}, {Samaras}, {Sanchez Gimenez}, {Sanna}, {Santove{\~n}a},
  {Sarasso}, {Schultheis}, {Sciacca}, {Segol}, {Segovia}, {S{\'e}gransan},
  {Semeux}, {Shahaf}, {Siddiqui}, {Siebert}, {Siltala}, {Slezak}, {Smart},
  {Solano}, {Solitro}, {Souami}, {Souchay}, {Spagna}, {Spoto}, {Steele},
  {Steidelm{\"u}ller}, {Stephenson}, {S{\"u}veges}, {Szabados}, {Szegedi-Elek},
  {Taris}, {Tauran}, {Taylor}, {Teixeira}, {Thuillot}, {Tonello}, {Torra},
  {Torra}, {Turon}, {Unger}, {Vaillant}, {van Dillen}, {Vanel}, {Vecchiato},
  {Viala}, {Vicente}, {Voutsinas}, {Weiler}, {Wevers}, {Wyrzykowski}, {Yoldas},
  {Yvard}, {Zhao}, {Zorec}, {Zucker}, {Zurbach}, \& {Zwitter}}]{gaiaedr3}
---. 2021, \aap, 649, A1

\bibitem[{{Gaia Collaboration} {et~al.}(2022{\natexlab{a}}){Gaia
  Collaboration}, {Drimmel}, {Romero-Gomez}, {Chemin}, {Ramos}, {Poggio},
  {Ripepi}, {Andrae}, {Blomme}, {Cantat-Gaudin}, {Castro-Ginard}, {Clementini},
  {Figueras}, {Fouesneau}, {Fremat}, {Jardine}, {Khanna}, {Lobel}, {Marshall},
  {Muraveva}, {Brown}, {Vallenari}, {Prusti}, {de Bruijne}, {Arenou},
  {Babusiaux}, {Biermann}, {Creevey}, {Ducourant}, {Evans}, {Eyer}, {Guerra},
  {Hutton}, {Jordi}, {Klioner}, {Lammers}, {Lindegren}, {Luri}, {Mignard},
  {Panem}, {Pourbaix}, {Randich}, {Sartoretti}, {Soubiran}, {Tanga}, {Walton},
  {Bailer-Jones}, {Bastian}, {Jansen}, {Katz}, {Lattanzi}, {van Leeuwen},
  {Bakker}, {Cacciari}, {Casta{\~n}eda}, {De Angeli}, {Fabricius}, {Galluccio},
  {Guerrier}, {Heiter}, {Masana}, {Messineo}, {Mowlavi}, {Nicolas},
  {Nienartowicz}, {Pailler}, {Panuzzo}, {Riclet}, {Roux}, {Seabroke},
  {Sordo{\o}rcit}, {Th{\'e}venin}, {Gracia-Abril}, {Portell}, {Teyssier},
  {Altmann}, {Audard}, {Bellas-Velidis}, {Benson}, {Berthier}, {Burgess},
  {Busonero}, {Busso}, {C{\'a}novas}, {Carry}, {Cellino}, {Cheek}, {Damerdji},
  {Davidson}, {de Teodoro}, {Nu{\~n}ez Campos}, {Delchambre}, {Dell'Oro},
  {Esquej}, {Fern{\'a}ndez-Hern{\'a}ndez}, {Fraile}, {Garabato},
  {Garc{\'\i}a-Lario}, {Gosset}, {Haigron}, {Halbwachs}, {Hambly}, {Harrison},
  {Hern{\'a}ndez}, {Hestroffer}, {Hodgkin}, {Holl}, {Jan{\ss}en}, {Jevardat de
  Fombelle}, {Jordan}, {Krone-Martins}, {Lanzafame}, {L{\"o}ffler}, {Marchal},
  {Marrese}, {Moitinho}, {Muinonen}, {Osborne}, {Pancino}, {Pauwels},
  {Recio-Blanco}, {Reyl{\'e}}, {Riello}, {Rimoldini}, {Roegiers}, {Rybizki},
  {Sarro}, {Siopis}, {Smith}, {Sozzetti}, {Utrilla}, {van Leeuwen}, {Abbas},
  {{\'A}brah{\'a}m}, {Abreu Aramburu}, {Aerts}, {Aguado}, {Ajaj},
  {Aldea-Montero}, {Altavilla}, {{\'A}lvarez}, {Alves}, {Anders}, {Anderson},
  {Anglada Varela}, {Antoja}, {Baines}, {Baker}, {Balaguer-N{\'u}{\~n}ez},
  {Balbinot}, {Balog}, {Barache}, {Barbato}, {Barros}, {Barstow},
  {Bartolom{\'e}}, {Bassilana}, {Bauchet}, {Becciani}, {Bellazzini},
  {Berihuete}, {Bernet}, {Bertone}, {Bianchi}, {Binnenfeld}, {Blanco-Cuaresma},
  {Blazere}, {Boch}, {Bombrun}, {Bossini}, {Bouquillon}, {Bragaglia},
  {Bramante}, {Breedt}, {Bressan}, {Brouillet}, {Brugaletta}, {Bucciarelli},
  {Burlacu}, {Butkevich}, {Buzzi}, {Caffau}, {Cancelliere}, {Carballo},
  {Carlucci}, {Carnerero}, {Carrasco}, {Casamiquela}, {Castellani}, {Chaoul},
  {Charlot}, {Chiaramida}, {Chiavassa}, {Chornay}, {Comoretto}, {Contursi},
  {Cooper}, {Cornez}, {Cowell}, {Crifo}, {Cropper}, {Crosta}, {Crowley},
  {Dafonte}, {Dapergolas}, {David}, {David}, {de Laverny}, {De Luise}, {De
  March}, {De Ridder}, {de Souza}, {de Torres}, {del Peloso}, {del Pozo},
  {Delbo}, {Delgado}, {Delisle}, {Demouchy}, {Dharmawardena}, {Di Matteo},
  {Diakite}, {Diener}, {Distefano}, {Dolding}, {Edvardsson}, {Enke}, {Fabre},
  {Fabrizio}, {Faigler}, {Fedorets}, {Fernique}, {Fienga}, {Fournier},
  {Fouron}, {Fragkoudi}, {Gai}, {Garcia-Gutierrez}, {Garcia-Reinaldos},
  {Garc{\'\i}a-Torres}, {Garofalo}, {Gavel}, {Gavras}, {Gerlach}, {Geyer},
  {Giacobbe}, {Gilmore}, {Girona}, {Giuffrida}, {Gomel}, {Gomez},
  {Gonz{\'a}lez-N{\'u}{\~n}ez}, {Gonz{\'a}lez-Santamar{\'\i}a},
  {Gonz{\'a}lez-Vidal}, {Granvik}, {Guillout}, {Guiraud},
  {Guti{\'e}rrez-S{\'a}nchez}, {Guy}, {Hatzidimitriou}, {Hauser}, {Haywood},
  {Helmer}, {Helmi}, {Sarmiento}, {Hidalgo}, {Hilger}, {H{\l}adczuk}, {Hobbs},
  {Holland}, {Huckle}, {Jasniewicz}, {Jean-Antoine Piccolo},
  {Jim{\'e}nez-Arranz}, {Jorissen}, {Juaristi Campillo}, {Julbe}, {Karbevska},
  {Kervella}, {Kontizas}, {Kordopatis}, {Korn}, {K{\'o}sp{\'a}l},
  {Kostrzewa-Rutkowska}, {Kruszy{\'n}ska}, {Kun}, {Laizeau}, {Lambert},
  {Lanza}, {Lasne}, {Le Campion}, {Lebreton}, {Lebzelter}, {Leccia}, {Leclerc},
  {Lecoeur-Taibi}, {Liao}, {Licata}, {Lindstr{\o}m}, {Lister}, {Livanou},
  {Lorca}, {Loup}, {Madrero Pardo}, {Magdaleno Romeo}, {Managau}, {Mann},
  {Manteiga}, {Marchant}, {Marconi}, {Marcos}, {Marcos Santos}, {Mar{\'\i}n
  Pina}, {Marinoni}, {Marocco}, {Polo}, {Mart{\'\i}n-Fleitas}, {Marton},
  {Mary}, {Masip}, {Massari}, {Mastrobuono-Battisti}, {Mazeh}, {McMillan},
  {Messina}, {Michalik}, {Millar}, {Mints}, {Molina}, {Molinaro}, {Moln{\'a}r},
  {Monari}, {Mongui{\'o}}, {Montegriffo}, {Montero}, {Mor}, {Mora},
  {Morbidelli}, {Morel}, {Morris}, {Murphy}, {Musella}, {Nagy}, {Noval},
  {Oca{\~n}a}, {Ogden}, {Ordenovic}, {Osinde}, {Pagani}, {Pagano}, {Palaversa},
  {Palicio}, {Pallas-Quintela}, {Panahi}, {Payne-Wardenaar}, {Pe{\~n}alosa
  Esteller}, {Penttil{\"a}}, {Pichon}, {Piersimoni}, {Pineau}, {Plachy},
  {Plum}, {Pr{\v{s}}a}, {Pulone}, {Racero}, {Ragaini}, {Rainer}, {Raiteri},
  {Rambaux}, {Ramos-Lerate}, {Re Fiorentin}, {Regibo}, {Richards}, {Rios Diaz},
  {Riva}, {Rix}, {Rixon}, {Robichon}, {Robin}, {Robin}, {Roelens}, {Rogues},
  {Rohrbasser}, {Rowell}, {Royer}, {Ruz Mieres}, {Rybicki}, {Sadowski},
  {S{\'a}ez N{\'u}{\~n}ez}, {Sagrist{\`a} Sell{\'e}s}, {Sahlmann}, {Salguero},
  {Samaras}, {Sanchez Gimenez}, {Sanna}, {Santove{\~n}a}, {Sarasso},
  {Schultheis}, {Sciacca}, {Segol}, {Segovia}, {S{\'e}gransan}, {Semeux},
  {Shahaf}, {Siddiqui}, {Siebert}, {Siltala}, {Silvelo}, {Slezak}, {Slezak},
  {Smart}, {Snaith}, {Solano}, {Solitro}, {Souami}, {Souchay}, {Spagna},
  {Spina}, {Spoto}, {Steele}, {Steidelm{\"u}ller}, {Stephenson}, {S{\"u}veges},
  {Surdej}, {Szabados}, {Szegedi-Elek}, {Taris}, {Taylo}, {Teixeira},
  {Tolomei}, {Tonello}, {Torra}, {Torra}, {Torralba Elipe}, {Trabucchi},
  {Tsounis}, {Turon}, {Ulla}, {Unger}, {Vaillant}, {van Dillen}, {van Reeven},
  {Vanel}, {Vecchiato}, {Viala}, {Vicente}, {Voutsinas}, {Weiler}, {Wevers},
  {Wyrzykowski}, {Yoldas}, {Yvard}, {Zhao}, {Zorec}, {Zucker}, \&
  {Zwitter}}]{gc20210}
{Gaia Collaboration}, {Drimmel}, R., {Romero-Gomez}, M., {et~al.}
  2022{\natexlab{a}}, arXiv e-prints, arXiv:2206.06207

\bibitem[{{Gaia Collaboration} {et~al.}(2022{\natexlab{b}}){Gaia
  Collaboration}, {Vallenari}, {Brown}, {Prusti}, {de Bruijne}, {Arenou},
  {Babusiaux}, {Biermann}, {Creevey}, {Ducourant}, {Evans}, {Eyer}, {Guerra},
  {Hutton}, {Jordi}, {Klioner}, {Lammers}, {Lindegren}, {Luri}, {Mignard},
  {Panem}, {Pourbaix}, {Randich}, {Sartoretti}, {Soubiran}, {Tanga}, {Walton},
  {Bailer-Jones}, {Bastian}, {Drimmel}, {Jansen}, {Katz}, {Lattanzi}, {van
  Leeuwen}, {Bakker}, {Cacciari}, {Casta{\~n}eda}, {De Angeli}, {Fabricius},
  {Fouesneau}, {Fr{\'e}mat}, {Galluccio}, {Guerrier}, {Heiter}, {Masana},
  {Messineo}, {Mowlavi}, {Nicolas}, {Nienartowicz}, {Pailler}, {Panuzzo},
  {Riclet}, {Roux}, {Seabroke}, {Sordo{\o}rcit}, {Th{\'e}venin},
  {Gracia-Abril}, {Portell}, {Teyssier}, {Altmann}, {Andrae}, {Audard},
  {Bellas-Velidis}, {Benson}, {Berthier}, {Blomme}, {Burgess}, {Busonero},
  {Busso}, {C{\'a}novas}, {Carry}, {Cellino}, {Cheek}, {Clementini},
  {Damerdji}, {Davidson}, {de Teodoro}, {Nu{\~n}ez Campos}, {Delchambre},
  {Dell'Oro}, {Esquej}, {Fern{\'a}ndez-Hern{\'a}ndez}, {Fraile}, {Garabato},
  {Garc{\'\i}a-Lario}, {Gosset}, {Haigron}, {Halbwachs}, {Hambly}, {Harrison},
  {Hern{\'a}ndez}, {Hestroffer}, {Hodgkin}, {Holl}, {Jan{\ss}en}, {Jevardat de
  Fombelle}, {Jordan}, {Krone-Martins}, {Lanzafame}, {L{\"o}ffler}, {Marchal},
  {Marrese}, {Moitinho}, {Muinonen}, {Osborne}, {Pancino}, {Pauwels},
  {Recio-Blanco}, {Reyl{\'e}}, {Riello}, {Rimoldini}, {Roegiers}, {Rybizki},
  {Sarro}, {Siopis}, {Smith}, {Sozzetti}, {Utrilla}, {van Leeuwen}, {Abbas},
  {{\'A}brah{\'a}m}, {Abreu Aramburu}, {Aerts}, {Aguado}, {Ajaj},
  {Aldea-Montero}, {Altavilla}, {{\'A}lvarez}, {Alves}, {Anders}, {Anderson},
  {Anglada Varela}, {Antoja}, {Baines}, {Baker}, {Balaguer-N{\'u}{\~n}ez},
  {Balbinot}, {Balog}, {Barache}, {Barbato}, {Barros}, {Barstow},
  {Bartolom{\'e}}, {Bassilana}, {Bauchet}, {Becciani}, {Bellazzini},
  {Berihuete}, {Bernet}, {Bertone}, {Bianchi}, {Binnenfeld}, {Blanco-Cuaresma},
  {Blazere}, {Boch}, {Bombrun}, {Bossini}, {Bouquillon}, {Bragaglia},
  {Bramante}, {Breedt}, {Bressan}, {Brouillet}, {Brugaletta}, {Bucciarelli},
  {Burlacu}, {Butkevich}, {Buzzi}, {Caffau}, {Cancelliere}, {Cantat-Gaudin},
  {Carballo}, {Carlucci}, {Carnerero}, {Carrasco}, {Casamiquela}, {Castellani},
  {Castro-Ginard}, {Chaoul}, {Charlot}, {Chemin}, {Chiaramida}, {Chiavassa},
  {Chornay}, {Comoretto}, {Contursi}, {Cooper}, {Cornez}, {Cowell}, {Crifo},
  {Cropper}, {Crosta}, {Crowley}, {Dafonte}, {Dapergolas}, {David}, {David},
  {de Laverny}, {De Luise}, {De March}, {De Ridder}, {de Souza}, {de Torres},
  {del Peloso}, {del Pozo}, {Delbo}, {Delgado}, {Delisle}, {Demouchy},
  {Dharmawardena}, {Di Matteo}, {Diakite}, {Diener}, {Distefano}, {Dolding},
  {Edvardsson}, {Enke}, {Fabre}, {Fabrizio}, {Faigler}, {Fedorets}, {Fernique},
  {Fienga}, {Figueras}, {Fournier}, {Fouron}, {Fragkoudi}, {Gai},
  {Garcia-Gutierrez}, {Garcia-Reinaldos}, {Garc{\'\i}a-Torres}, {Garofalo},
  {Gavel}, {Gavras}, {Gerlach}, {Geyer}, {Giacobbe}, {Gilmore}, {Girona},
  {Giuffrida}, {Gomel}, {Gomez}, {Gonz{\'a}lez-N{\'u}{\~n}ez},
  {Gonz{\'a}lez-Santamar{\'\i}a}, {Gonz{\'a}lez-Vidal}, {Granvik}, {Guillout},
  {Guiraud}, {Guti{\'e}rrez-S{\'a}nchez}, {Guy}, {Hatzidimitriou}, {Hauser},
  {Haywood}, {Helmer}, {Helmi}, {Sarmiento}, {Hidalgo}, {Hilger},
  {H{\l}adczuk}, {Hobbs}, {Holland}, {Huckle}, {Jardine}, {Jasniewicz},
  {Jean-Antoine Piccolo}, {Jim{\'e}nez-Arranz}, {Jorissen}, {Juaristi
  Campillo}, {Julbe}, {Karbevska}, {Kervella}, {Khanna}, {Kontizas},
  {Kordopatis}, {Korn}, {K{\'o}sp{\'a}l}, {Kostrzewa-Rutkowska},
  {Kruszy{\'n}ska}, {Kun}, {Laizeau}, {Lambert}, {Lanza}, {Lasne}, {Le
  Campion}, {Lebreton}, {Lebzelter}, {Leccia}, {Leclerc}, {Lecoeur-Taibi},
  {Liao}, {Licata}, {Lindstr{\o}m}, {Lister}, {Livanou}, {Lobel}, {Lorca},
  {Loup}, {Madrero Pardo}, {Magdaleno Romeo}, {Managau}, {Mann}, {Manteiga},
  {Marchant}, {Marconi}, {Marcos}, {Marcos Santos}, {Mar{\'\i}n Pina},
  {Marinoni}, {Marocco}, {Marshall}, {Polo}, {Mart{\'\i}n-Fleitas}, {Marton},
  {Mary}, {Masip}, {Massari}, {Mastrobuono-Battisti}, {Mazeh}, {McMillan},
  {Messina}, {Michalik}, {Millar}, {Mints}, {Molina}, {Molinaro}, {Moln{\'a}r},
  {Monari}, {Mongui{\'o}}, {Montegriffo}, {Montero}, {Mor}, {Mora},
  {Morbidelli}, {Morel}, {Morris}, {Muraveva}, {Murphy}, {Musella}, {Nagy},
  {Noval}, {Oca{\~n}a}, {Ogden}, {Ordenovic}, {Osinde}, {Pagani}, {Pagano},
  {Palaversa}, {Palicio}, {Pallas-Quintela}, {Panahi}, {Payne-Wardenaar},
  {Pe{\~n}alosa Esteller}, {Penttil{\"a}}, {Pichon}, {Piersimoni}, {Pineau},
  {Plachy}, {Plum}, {Poggio}, {Pr{\v{s}}a}, {Pulone}, {Racero}, {Ragaini},
  {Rainer}, {Raiteri}, {Rambaux}, {Ramos}, {Ramos-Lerate}, {Re Fiorentin},
  {Regibo}, {Richards}, {Rios Diaz}, {Ripepi}, {Riva}, {Rix}, {Rixon},
  {Robichon}, {Robin}, {Robin}, {Roelens}, {Rogues}, {Rohrbasser},
  {Romero-G{\'o}mez}, {Rowell}, {Royer}, {Ruz Mieres}, {Rybicki}, {Sadowski},
  {S{\'a}ez N{\'u}{\~n}ez}, {Sagrist{\`a} Sell{\'e}s}, {Sahlmann}, {Salguero},
  {Samaras}, {Sanchez Gimenez}, {Sanna}, {Santove{\~n}a}, {Sarasso},
  {Schultheis}, {Sciacca}, {Segol}, {Segovia}, {S{\'e}gransan}, {Semeux},
  {Shahaf}, {Siddiqui}, {Siebert}, {Siltala}, {Silvelo}, {Slezak}, {Slezak},
  {Smart}, {Snaith}, {Solano}, {Solitro}, {Souami}, {Souchay}, {Spagna},
  {Spina}, {Spoto}, {Steele}, {Steidelm{\"u}ller}, {Stephenson}, {S{\"u}veges},
  {Surdej}, {Szabados}, {Szegedi-Elek}, {Taris}, {Taylo}, {Teixeira},
  {Tolomei}, {Tonello}, {Torra}, {Torra}, {Torralba Elipe}, {Trabucchi},
  {Tsounis}, {Turon}, {Ulla}, {Unger}, {Vaillant}, {van Dillen}, {van Reeven},
  {Vanel}, {Vecchiato}, {Viala}, {Vicente}, {Voutsinas}, {Weiler}, {Wevers},
  {Wyrzykowski}, {Yoldas}, {Yvard}, {Zhao}, {Zorec}, {Zucker}, \&
  {Zwitter}}]{gaiadr3}
{Gaia Collaboration}, {Vallenari}, A., {Brown}, A.~G.~A., {et~al.}
  2022{\natexlab{b}}, arXiv e-prints, arXiv:2208.00211

\bibitem[{{Gum} {et~al.}(1960){Gum}, {Kerr}, \& {Westerhout}}]{Gum60}
{Gum}, C.~S., {Kerr}, F.~J., \& {Westerhout}, G. 1960, \mnras, 121, 132

\bibitem[{{He} {et~al.}(2023{\natexlab{a}}){He}, {Liu}, {Luo}, {Wang}, \&
  {Jiang}}]{he23a}
{He}, Z., {Liu}, X., {Luo}, Y., {Wang}, K., \& {Jiang}, Q. 2023{\natexlab{a}},
  \apjs, 264, 8

\bibitem[{{He} {et~al.}(2023{\natexlab{b}}){He}, {Luo}, {Wang}, {Ren}, {Peng},
  {Cui}, {Liu}, \& {Jiang}}]{he23b}
{He}, Z., {Luo}, Y., {Wang}, K., {et~al.} 2023{\natexlab{b}}, arXiv e-prints,
  arXiv:2305.10269

\bibitem[{{He} {et~al.}(2022){He}, {Li}, {Zhong}, {Liu}, {Bai}, {Qin}, {Jiang},
  {Zhang}, \& {Chen}}]{he22a}
{He}, Z., {Li}, C., {Zhong}, J., {et~al.} 2022, \apjs, 260, 8

\bibitem[{{He} {et~al.}(2021){He}, {Xu}, {Hao}, {Wu}, \& {Li}}]{he21}
{He}, Z.-H., {Xu}, Y., {Hao}, C.-J., {Wu}, Z.-Y., \& {Li}, J.-J. 2021, Research
  in Astronomy and Astrophysics, 21, 093

\bibitem[{{Hunter} \& {Toomre}(1969)}]{Hunter69}
{Hunter}, C., \& {Toomre}, A. 1969, \apj, 155, 747

\bibitem[{{Jarrett} {et~al.}(2003){Jarrett}, {Chester}, {Cutri}, {Schneider},
  \& {Huchra}}]{Jarrett03}
{Jarrett}, T.~H., {Chester}, T., {Cutri}, R., {Schneider}, S.~E., \& {Huchra},
  J.~P. 2003, \aj, 125, 525

\bibitem[{{Kerr}(1957)}]{Kerr57}
{Kerr}, F.~J. 1957, \aj, 62, 93

\bibitem[{{Lemasle} {et~al.}(2022){Lemasle}, {Lala}, {Kovtyukh}, {Hanke},
  {Prudil}, {Bono}, {Braga}, {da Silva}, {Fabrizio}, {Fiorentino},
  {Fran{\c{c}}ois}, {Grebel}, \& {Kniazev}}]{Lemasle22}
{Lemasle}, B., {Lala}, H.~N., {Kovtyukh}, V., {et~al.} 2022, \aap, 668, A40

\bibitem[{{Levine} {et~al.}(2006){Levine}, {Blitz}, \& {Heiles}}]{Levine06}
{Levine}, E.~S., {Blitz}, L., \& {Heiles}, C. 2006, \apj, 643, 881

\bibitem[{{Li} {et~al.}(2023){Li}, {Wang}, {Luo}, {L{\'o}pez-Corredoira},
  {Ting}, \& {Chrob{\'a}kov{\'a}}}]{Li23}
{Li}, X., {Wang}, H.-F., {Luo}, Y.-P., {et~al.} 2023, \apj, 943, 88

\bibitem[{{Li} {et~al.}(2020){Li}, {Huang}, {Chen}, {Wang}, {Sun}, {Guo}, {Li},
  \& {Liu}}]{Li20}
{Li}, X.~Y., {Huang}, Y., {Chen}, B.~Q., {et~al.} 2020, \apj, 901, 56

\bibitem[{{Lindegren}(2020)}]{Lindegren20}
{Lindegren}, L. 2020, \aap, 633, A1

\bibitem[{{Lindegren} {et~al.}(2018){Lindegren}, {Hern{\'a}ndez}, {Bombrun},
  {Klioner}, {Bastian}, {Ramos-Lerate}, {de Torres}, {Steidelm{\"u}ller},
  {Stephenson}, {Hobbs}, {Lammers}, {Biermann}, {Geyer}, {Hilger}, {Michalik},
  {Stampa}, {McMillan}, {Casta{\~n}eda}, {Clotet}, {Comoretto}, {Davidson},
  {Fabricius}, {Gracia}, {Hambly}, {Hutton}, {Mora}, {Portell}, {van Leeuwen},
  {Abbas}, {Abreu}, {Altmann}, {Andrei}, {Anglada}, {Balaguer-N{\'u}{\~n}ez},
  {Barache}, {Becciani}, {Bertone}, {Bianchi}, {Bouquillon}, {Bourda},
  {Br{\"u}semeister}, {Bucciarelli}, {Busonero}, {Buzzi}, {Cancelliere},
  {Carlucci}, {Charlot}, {Cheek}, {Crosta}, {Crowley}, {de Bruijne}, {de
  Felice}, {Drimmel}, {Esquej}, {Fienga}, {Fraile}, {Gai}, {Garralda},
  {Gonz{\'a}lez-Vidal}, {Guerra}, {Hauser}, {Hofmann}, {Holl}, {Jordan},
  {Lattanzi}, {Lenhardt}, {Liao}, {Licata}, {Lister}, {L{\"o}ffler},
  {Marchant}, {Martin-Fleitas}, {Messineo}, {Mignard}, {Morbidelli}, {Poggio},
  {Riva}, {Rowell}, {Salguero}, {Sarasso}, {Sciacca}, {Siddiqui}, {Smart},
  {Spagna}, {Steele}, {Taris}, {Torra}, {van Elteren}, {van Reeven}, \&
  {Vecchiato}}]{Lindegren18}
{Lindegren}, L., {Hern{\'a}ndez}, J., {Bombrun}, A., {et~al.} 2018, \aap, 616,
  A2

\bibitem[{{Lindegren} {et~al.}(2021){Lindegren}, {Bastian}, {Biermann},
  {Bombrun}, {de Torres}, {Gerlach}, {Geyer}, {Hern{\'a}ndez}, {Hilger},
  {Hobbs}, {Klioner}, {Lammers}, {McMillan}, {Ramos-Lerate},
  {Steidelm{\"u}ller}, {Stephenson}, \& {van Leeuwen}}]{Lindegren21}
{Lindegren}, L., {Bastian}, U., {Biermann}, M., {et~al.} 2021, \aap, 649, A4

\bibitem[{{Liu} {et~al.}(2017){Liu}, {Tian}, \& {Wan}}]{Liu17}
{Liu}, C., {Tian}, H.-J., \& {Wan}, J.-C. 2017, in Formation and Evolution of
  Galaxy Outskirts, ed. A.~{Gil de Paz}, J.~H. {Knapen}, \& J.~C. {Lee}, Vol.
  321, 6--9

\bibitem[{{L{\'o}pez-Corredoira} {et~al.}(2014){L{\'o}pez-Corredoira}, {Abedi},
  {Garz{\'o}n}, \& {Figueras}}]{Corredoira14}
{L{\'o}pez-Corredoira}, M., {Abedi}, H., {Garz{\'o}n}, F., \& {Figueras}, F.
  2014, \aap, 572, A101

\bibitem[{{L{\'o}pez-Corredoira} {et~al.}(2002){L{\'o}pez-Corredoira},
  {Cabrera-Lavers}, {Garz{\'o}n}, \& {Hammersley}}]{Corredoira02}
{L{\'o}pez-Corredoira}, M., {Cabrera-Lavers}, A., {Garz{\'o}n}, F., \&
  {Hammersley}, P.~L. 2002, \aap, 394, 883

\bibitem[{{Mermilliod} {et~al.}(2009){Mermilliod}, {Mayor}, \&
  {Udry}}]{Mermilliod09}
{Mermilliod}, J.~C., {Mayor}, M., \& {Udry}, S. 2009, \aap, 498, 949

\bibitem[{{Miyamoto} \& {Zhu}(1998)}]{Miyamoto98}
{Miyamoto}, M., \& {Zhu}, Z. 1998, \aj, 115, 1483

\bibitem[{{Nakanishi} \& {Sofue}(2003)}]{Nakanishi03}
{Nakanishi}, H., \& {Sofue}, Y. 2003, \pasj, 55, 191

\bibitem[{{Poggio} {et~al.}(2020){Poggio}, {Drimmel}, {Andrae}, {Bailer-Jones},
  {Fouesneau}, {Lattanzi}, {Smart}, \& {Spagna}}]{Poggio20}
{Poggio}, E., {Drimmel}, R., {Andrae}, R., {et~al.} 2020, Nature Astronomy, 4,
  590

\bibitem[{{Poggio} {et~al.}(2017){Poggio}, {Drimmel}, {Smart}, {Spagna}, \&
  {Lattanzi}}]{Poggio17}
{Poggio}, E., {Drimmel}, R., {Smart}, R.~L., {Spagna}, A., \& {Lattanzi}, M.~G.
  2017, \aap, 601, A115

\bibitem[{{Poggio} {et~al.}(2018){Poggio}, {Drimmel}, {Lattanzi}, {Smart},
  {Spagna}, {Andrae}, {Bailer-Jones}, {Fouesneau}, {Antoja}, {Babusiaux},
  {Evans}, {Figueras}, {Katz}, {Reyl{\'e}}, {Robin}, {Romero-G{\'o}mez}, \&
  {Seabroke}}]{Poggio18}
{Poggio}, E., {Drimmel}, R., {Lattanzi}, M.~G., {et~al.} 2018, \mnras, 481, L21

\bibitem[{{Pringle}(1992)}]{Pringle92}
{Pringle}, J.~E. 1992, \mnras, 258, 811

\bibitem[{{Quinn} {et~al.}(1993){Quinn}, {Hernquist}, \& {Fullagar}}]{Quinn93}
{Quinn}, P.~J., {Hernquist}, L., \& {Fullagar}, D.~P. 1993, \apj, 403, 74

\bibitem[{{Reid} {et~al.}(2019){Reid}, {Menten}, {Brunthaler}, {Zheng}, {Dame},
  {Xu}, {Li}, {Sakai}, {Wu}, {Immer}, {Zhang}, {Sanna}, {Moscadelli}, {Rygl},
  {Bartkiewicz}, {Hu}, {Quiroga-Nu{\~n}ez}, \& {van Langevelde}}]{Reid19}
{Reid}, M.~J., {Menten}, K.~M., {Brunthaler}, A., {et~al.} 2019, \apj, 885, 131

\bibitem[{{Robin} {et~al.}(2003){Robin}, {Reyl{\'e}}, {Derri{\`e}re}, \&
  {Picaud}}]{Robin03}
{Robin}, A.~C., {Reyl{\'e}}, C., {Derri{\`e}re}, S., \& {Picaud}, S. 2003,
  \aap, 409, 523

\bibitem[{{Romero-G{\'o}mez} {et~al.}(2019){Romero-G{\'o}mez}, {Mateu},
  {Aguilar}, {Figueras}, \& {Castro-Ginard}}]{Romero19}
{Romero-G{\'o}mez}, M., {Mateu}, C., {Aguilar}, L., {Figueras}, F., \&
  {Castro-Ginard}, A. 2019, \aap, 627, A150

\bibitem[{{Ro{\v{s}}kar} {et~al.}(2010){Ro{\v{s}}kar}, {Debattista}, {Brooks},
  {Quinn}, {Brook}, {Governato}, {Dalcanton}, \& {Wadsley}}]{Rok10}
{Ro{\v{s}}kar}, R., {Debattista}, V.~P., {Brooks}, A.~M., {et~al.} 2010,
  \mnras, 408, 783

\bibitem[{{Sancisi}(1976)}]{Sancisi76}
{Sancisi}, R. 1976, \aap, 53, 159

\bibitem[{{Semczuk} {et~al.}(2023){Semczuk}, {Dehnen}, {Sch{\"o}nrich}, \&
  {Athanassoula}}]{Semczuk23}
{Semczuk}, M., {Dehnen}, W., {Sch{\"o}nrich}, R., \& {Athanassoula}, E. 2023,
  \mnras, 519, 902

\bibitem[{{Shen} \& {Sellwood}(2006)}]{Shen06}
{Shen}, J., \& {Sellwood}, J.~A. 2006, \mnras, 370, 2

\bibitem[{{Skowron} {et~al.}(2019{\natexlab{a}}){Skowron}, {Skowron},
  {Mr{\'o}z}, {Udalski}, {Pietrukowicz}, {Soszy{\'n}ski}, {Szyma{\'n}ski},
  {Poleski}, {Koz{\l}owski}, {Ulaczyk}, {Rybicki}, \& {Iwanek}}]{Skowron19}
{Skowron}, D.~M., {Skowron}, J., {Mr{\'o}z}, P., {et~al.} 2019{\natexlab{a}},
  Science, 365, 478

\bibitem[{{Skowron} {et~al.}(2019{\natexlab{b}}){Skowron}, {Skowron},
  {Mr{\'o}z}, {Udalski}, {Pietrukowicz}, {Soszy{\'n}ski}, {Szyma{\'n}ski},
  {Poleski}, {Koz{\l}owski}, {Ulaczyk}, {Rybicki}, {Iwanek}, {. Wrona}, \&
  {Gromadzki}}]{Skowron19b}
---. 2019{\natexlab{b}}, \actaa, 69, 305

\bibitem[{{Sparke} \& {Casertano}(1988)}]{Sparke88}
{Sparke}, L.~S., \& {Casertano}, S. 1988, \mnras, 234, 873

\bibitem[{{Sun} {et~al.}(2015){Sun}, {Xu}, {Yang}, {Li}, {Du}, {Zhang}, \&
  {Zhou}}]{Sun15}
{Sun}, Y., {Xu}, Y., {Yang}, J., {et~al.} 2015, \apjl, 798, L27

\bibitem[{{Thilker} {et~al.}(2005){Thilker}, {Bianchi}, {Boissier}, {Gil de
  Paz}, {Madore}, {Martin}, {Meurer}, {Neff}, {Rich}, {Schiminovich},
  {Seibert}, {Wyder}, {Barlow}, {Byun}, {Donas}, {Forster}, {Friedman},
  {Heckman}, {Jelinsky}, {Lee}, {Malina}, {Milliard}, {Morrissey}, {Siegmund},
  {Small}, {Szalay}, \& {Welsh}}]{Thilker05}
{Thilker}, D.~A., {Bianchi}, L., {Boissier}, S., {et~al.} 2005, \apjl, 619, L79

\bibitem[{{Velazquez} \& {White}(1999)}]{Velazquez99}
{Velazquez}, H., \& {White}, S. D.~M. 1999, \mnras, 304, 254

\bibitem[{{Voskes} \& {Butler Burton}(2006)}]{Voskes06}
{Voskes}, T., \& {Butler Burton}, W. 2006, arXiv e-prints, astro

\bibitem[{{Wang} {et~al.}(2020){Wang}, {L{\'o}pez-Corredoira}, {Huang},
  {Chang}, {Zhang}, {Carlin}, {Chen}, {Chrob{\'a}kov{\'a}}, \& {Chen}}]{Wang20}
{Wang}, H.~F., {L{\'o}pez-Corredoira}, M., {Huang}, Y., {et~al.} 2020, \apj,
  897, 119

\bibitem[{{Zhu}(2000)}]{Zhu00}
{Zhu}, Z. 2000, \apss, 271, 353

\end{thebibliography}

\end{CJK*}

\end{document}